\documentclass[12pt,preprint]{aastex}
\newcommand\be{\begin{equation}}
\newcommand\ee{\end{equation}}
\newcommand\ba{\begin{eqnarray}}
\newcommand\ea{\end{eqnarray}}
\newcommand\eq{\begin{equation}}
\newcommand\en{\end{equation}}

\begin{document}
\title{Uncertainties in the S-Z selected cluster angular power spectrum}
\author{J. D. Cohn${}^{1}$, Kenji Kadota${}^{2,3}$}
\affil{
${}^1$ Space Sciences Lab and  Theoretical Astrophysics Center,
${}^2$ Department of Physics,\\
University of California, Berkeley, CA 94720, USA \\
${}^3$NASA/Fermilab Astrophysics Center, Fermi National Accelerator
Laboratory, Batavia, IL 60510 \\
{\rm email: ~jcohn@astron.berkeley.edu, kadota@fnal.gov}}
\begin{abstract}
Large SZ selected galaxy cluster surveys are beginning imminently. We compare 
the dependence of the galaxy cluster angular power spectrum
on cosmological parameters, different modeling assumptions and
statistical observational errors. We quantify the degeneracies between
theoretical assumptions such as the mass function and cosmological
parameters such as $\sigma_8$. 
We also identify a rough scaling behavior of this angular power spectrum
with $\sigma_8$ alone.

\end{abstract}
\section{Introduction}
The Sunyaev Zel'dovich (hereafter SZ) effect is the upscattering of cosmic
microwave background photons by the hot electrons in galaxy clusters
(Sunyaev \& Zel'dovich \cite{SunZel72,SunZel80}).
As the surface brightness of a cluster in the SZ is redshift independent, 
and the signal is relatively insensitive to the
poorly known cluster core structure,
the SZ effect is a powerful tool to select and study galaxy clusters.
The field of SZ measurements has progressed rapidly from a handful of 
SZ detections for already-known galaxy clusters to SZ maps of 
several clusters to the current stage: cluster surveys 
designed exclusively for SZ detection.  For reviews, see for
example Birkinshaw \cite{Bir99} and 
Carlstrom, Holder \& Reese \cite{CarHolRee02}.  
Some surveys are already in progress such as 
ACBAR\footnote{http://astronomy.sussex.ac.uk/$\sim$romer/research/blind.html}
and 
SZA\footnote{ http://astro.uchicago.edu/sza/overview.html}.  
Others are on the verge of taking data, such as
AMI\footnote{http://www.mrao.cam.ac.uk/telescopes/ami/index.html} and
APEX\footnote{http://bolo.berkeley.edu/apexsz/index.html},
and ACT\footnote{http://www.hep.upenn.edu/$\sim$angelica/act/act.html}
expects to be ready by the end of 2006.    Thousands of previously
unknown clusters will be observed by these experiments, starting with
APEX.  
Future surveys such as
SPT\footnote{http://astro.uchicago.edu/spt} and 
Planck\footnote{http://astro.estec.esa.nl/SA-general/Projects/Planck} are
taking place within the next five years and should 
observe tens of thousands of clusters.
For a full list of current and upcoming experiments, see for example the
CMB experiment page at LAMBDA\footnote{http://lambda.gsfc.nasa.gov/}.

The prospect of combining these theoretical predictions and 
observational data on galaxy clusters to constrain fundamental 
cosmological parameters has led to much excitement (e.g.
Holder, Haiman \& Mohr \cite{HolHaiMoh01}, Majumdar \& Mohr \cite{MajMoh03}
and references below).
With large survey data on the horizon, and the detailed design of
later experiments taking shape, it is timely to identify which
theoretical assumptions are the most crucial to pin down in
order to use the science harvest from these experiments.
The initial galaxy cluster information these surveys will produce
will be number counts and the angular correlation function/cluster power spectrum. 

We consider here the angular power spectrum of clusters found by an SZ 
survey.  We calculate the effect of cosmological, modeling and 
observational parameters on this power spectrum.  The cosmological parameters 
we consider are $\Omega_m$ and $\sigma_8$.  In the modeling,
analytic predictions (based on the extremely good fit of 
analytic models to dark matter simulations) can predict
the expected galaxy cluster counts, masses, and positions.  These 
fits often differ at about the 10\% level.  
In addition, cluster observations generally measure some proxy for the 
cluster mass (temperature, X-ray flux, SZ flux, velocities, shear in images).
These proxies generally depend on more complex astrophysical 
properties (gasdynamics, for instance, or the state of relaxation of 
the cluster), and/or are difficult to connect cleanly to the mass 
(e.g. weak lensing masses, Metzler, White \& Loken \cite{MetWhiLok01}).  
We consider different analytic fits and assumptions for mass proxies. 
For the observational parameters
we consider sensitivity and area of survey,
using those of APEX for illustration.
We quantify and compare the effects on the power spectrum 
of these uncertainties.  Many of 
these uncertainties are degenerate, we identify which parameter
changes are roughly equivalent, and which are not.

The SZ properties of clusters in upcoming surveys are of great interest
and many aspects have been studied in previous work.
Constraints from the angular correlations of
clusters found in the SZ have been studied by
Moscardini et al \cite{Mos02},
Diaferio et al \cite{Dia03},  
Mei \& Bartlett \cite{MeiBar03,MeiBar04}.  These papers each varied
a combination of different quantities (e.g. $\sigma_8$, normalization 
of SZ flux as function of mass, biases, scaling of SZ flux with mass
and redshift). Majumdar \& Mohr \cite{MajMoh03a} consider the two-dimensional
power spectrum and 
Wang et al \cite{Wan04} consider the three dimensional power spectrum
in conjunction with other measurements.
A similar quantity, the angular correlation of the SZ temperature
spectrum, has been considered by
Komatsu \&,  Kitayama \cite{KomKit99},  Komatsu \& Seljak 
\cite{KomSel02}, and Majumdar \& Mohr \cite{MohMaj03}. 
Battye \& Weller \cite{BatWel03} studied similar systematics to those
considered here, for number counts as a function of redshift, $dN/dz$.
Uncertainties which were treated separately in (different) previous works
can be degenerate, leading to potentially misleading interpretations.
We quantify the changes in cosmological parameters which mimic these
other theoretical assumptions.
 
In \S 2, we discuss the selection requirement for the 
SZ cluster catalogue, \S 3 defines the
angular power spectrum and angular correlation function.  Our new
results are in \S 4, which
considers the cosmological, modeling and observational 
uncertainties in the power spectrum and compares them.  
We discuss many of the uncertainties in detail, and then compare
the changes caused by varying many different assumptions.  We
both find degeneracies and find which assumptions
are not degenerate.  Some of our assumptions have been 
considered separately in other works mentioned above (often for
the correlation function rather than the power spectrum). 
By combining previously considered changes in assumptions and
newer ones all together in a homogeneous manner and comparing to the same 
standard reference model we are
able to quantify the relative importance of different
theoretical assumptions in relation to cosmological parameters and
to each other.  
\S 5 concludes.
Unless otherwise stated, we will take the Hubble constant to be
$h=0.7$, baryon fraction $\Omega_b h^2 = 0.02$, in a 
flat $\Lambda$CDM universe with $\Omega_m = 0.3, \Omega_\Lambda = 0.7$.

\section{SZ selected galaxy cluster catalogues}
An SZ selected galaxy cluster catalogue is one that includes all clusters
above a certain minimum SZ flux or (equivalently) above some minimum
$Y$ parameter $Y_{min}$.  (The $Y$ parameter is defined in detail below.)
For a general review of the SZ effect, see for example 
Birkinshaw \cite{Bir99}, Rephaeli \cite{Rep95}, and
Sunyaev \& Zel'dovich \cite{SunZel80}.
There are two SZ effects, thermal and kinetic.
The (frequency dependent) thermal SZ effect is the 
change in the CMB spectrum 
due to random thermal motion of the intracluster electrons, and the 
(frequency independent) kinetic SZ effect 
is the change in the CMB spectrum due to bulk peculiar motions.
The kinetic SZ effect is negligible compared to the thermal SZ effect 
for the purpose of cluster
selection under study here; consequently we restrict our
attention to the thermal SZ effect; 
``SZ effect'' hereafter means thermal SZ effect.

The thermal SZ effect can be described in terms of a CMB flux
increment or decrement using the dimensionless
Comptonization parameter $y(\theta)$,
\eq
\frac{\Delta T}{T_{CMB}} =  g(x)  y(\theta)
\en
where the prefactor $g(x) =  ( x \frac{e^x +1}{e^x -1} -4 )(1 + \delta_{rel}(x))$
includes frequency dependence,
$x = h_p \nu/(k_BT_{CMB}) = \nu/56.84 GHz$, and we will neglect
the small relativistic correction $\delta_{rel}$, which gives
only a few percent effect for the hottest clusters 
(Rephaeli \cite{Rep95a}, Itoh et al \cite{Itoetal98}, 
Nozawa et al. \cite{Nozetal00}, Fan \& Wu \cite{FanWu03}).
We will be interested in the integrated $Y$ parameter,
\eq
Y \equiv \int d\Omega y(\theta)
\en 
where the integral is over the solid angle subtended by the cluster.
This can be written in terms of cluster properties as
$y(\theta)$ is the integration of pressure
along the line of sight which passes at an angle $\theta$ away from
the center of the cluster:
\eq
y(\theta) = \frac{k_B \sigma_T}{m_e c^2 } \int d \ell n_e T_e \; .
\en
Here $h_p,\sigma_T,k_B, m_e,n_e,T_e$ are
respectively the Planck constant, Thomson cross section, Boltzmann constant,
electron mass, intracluster electron density and temperature.
The number of electrons along the line of sight to
the cluster mass is taken to be
\eq
d_A^2 \int d \Omega d \ell n_e = \frac{M_{vir}f_{gas}}{\mu_e m_p}
\en
where $d_A = \frac{1}{1+z} \frac{c}{H_0} \int_0^z \frac{1}{E(z')} dz'$, 
$E(z) = \sqrt{\Omega_m (1+z)^3 + \Omega_\Lambda}$,
$M_{vir}$ is the virial mass, $f_{gas}$ is the 
intra cluster gas fraction\footnote{Note that for a 
$M_{vir} = 10^{14} h^{-1} M_\odot$
cluster we get $f_{gas} = 0.06 h^{-3/2}$.  The factor of
$h$ is included to make connection with other definitions, as
Lin et al \cite{LinMohSta03} point out, the variation with $h$ is not
actually a simple scaling.  We keep $h$ fixed in this paper.
Note that measurements of $f_{gas}$ implicitly require 
gas physics theoretical modeling.} 
\eq
f_{gas} =
0.10 h^{-3/2} M_{15}^{0.148}/(1 + 0.10 M_{15}^{-0.25}) \;
\en 
(Lin, Mohr
and Stanford \cite{LinMohSta03},
$M_{15}=\frac{M_{vir}}{10^{15} h^{-1} M_\odot}$),
$\mu_e = 1.143$ is 
the mean mass per electron and $m_p$ is the proton mass.
Then the electron density weighted average temperature 
\eq
\langle T_e \rangle_n = \frac{\int d \ell n_e T_e}{\int d\ell n_e}
\en
is given, using virialization arguments (see e.g. Battye \& Weller
\cite{BatWel03}) by
\eq
\label{tstardef}
\langle T_e \rangle_n = T_{*} \left(\frac{M_{vir}}{10^{15} h^{-1} M_\odot}\right)^{2/3} 
\left(\Delta_c E(z)^2\right)^{1/3} \left(1 - 2 \frac{\Omega_\Lambda}{\Delta_c}\right)
\en
where $\Delta_c(z) = 18 \pi^2 + 82 (\Omega_m(z)-1) - 39 (\Omega_m(z)-1)^2$.
We have taken the $z$ dependence from Pierpaoli et al \cite{Pie02}.
Note this assumes that only electrons within the virial radius
are contributing to the SZ effect.

Although this form and a specific $T_*$ can be ``derived'' 
using virialization arguments, one
can also just define $T_*$ as the constant of proportionality in the
above.  The above mass-temperature relation seems to work well for X-ray
temperatures of high mass clusters and most measurements of $T_*$ for the 
above relation are done in the X-ray, we will call the X-ray value
$T_*^{X-ray}$.  Simulations find $T_*^{X-ray}$ to 
be $\sim 1.2$ keV, while observations tend to prefer a higher values, 
$T_*^{X-ray}$  $\sim 2.0 $ keV (for a recent compilation see, e.g., 
Huterer \& White \cite{HutWhi02}, and for detailed discussion of
subtleties in X-ray temperature definitions see, e.g. Mathiesen \& Evrard
\cite{MatEvr01} and more recently Borgani et al \cite{Bor04},
Mazzotta et al \cite{Mazetal04} and Rasia et al \cite{Rasetal04} and 
references therein; convergence is improving steadily (Kravtsov \cite{Kra05})
). 
For the calculations here we will need $T_*^{SZ}$.
There is no a priori reason 
why the X-ray temperature and the SZ temperature normalizations are
identical, as they get the bulk of their signal from different 
parts of the cluster.  We discuss this in more detail later in 
section \S 4.2 when we consider modeling uncertainties.
In addition, the above is in terms of
$M_{vir}$; several mass definitions are in use in the literature.  If
these differences are not taken into account correctly (White \cite{Whi01}) via
mass conversion, an apparent (but incorrect) change in $T_*^{SZ}$
will result, more discussion on this issue is in section \S 4.2.
Combining these definitions and using again
$M_{15}=\frac{M_{vir}}{10^{15} h^{-1} M_\odot}$,
we get
\eq
\label{ydef}
\begin{array}{ll}
Y &= \int d \Omega \frac{k_B \sigma_T}{m_e c^2 } \int d \ell n_e T_e \\
&= \frac{k_B \sigma_T}{m_e c^2 } \frac{f_{gas}M_{vir}}{\mu_e m_p}
\frac{1}{d_A^2} T^{SZ}_* M_{15}^{2/3} 
(\Delta_c E(z)^2)^{1/3} (1 - 2 \frac{\Omega_\Lambda}{\Delta_c})
\\
&= 1.69 \times 10^{3}  f_{gas} h \frac{T^{SZ}_*}{keV} 
M_{15}^{5/3} (\Delta_c E(z)^2)^{1/3} 
(1 - 2 \frac{\Omega_\Lambda}{\Delta_c})
(\frac{h^{-1} Mpc}{d_A})^{2} {\rm arcmin}^2  \\
\end{array}
\en
The resulting 
SZ effect is a small distortion of the CMB of order $ \sim 1 mK$. 
Results are often quoted in terms of flux,
with a conversion
\eq
F_\nu =  2.28 \times 10^4 
\frac{x^4 e^x }{(e^x -1)^2} \left( x \frac{e^x +1}{e^x -1} -4 \right)
\frac{Y}{{\rm arcmin}^2} mJy
\en
For 143 GHz, the $x$ dependent factor $\sim -4$
(which translates into $Y = -\Delta T/T$ for the $Y$ parameter),
for 90 GHz it is -3.3 and for 265 GHz it is +3.4.
The SZ effect switches from a decrement to an increment in the
CMB spectrum at 218 GHz.  Thus, one way of distinguishing the
thermal SZ effect from other sources, such as primary anisotropy or noise,
is to see if it changes at 218 GHz.

Specifics in going from this flux or corresponding $Y$ value to a 
cluster detection depend upon the particulars of each experiment.  
We will consider the idealized case where an SZ experiment will detect 
all clusters above some minimum $Y$ value, $Y_{min}$.  Experiment-specific 
analysis and followup will be necessary to make reliable cluster 
identification.\footnote{One issue is the effect of beam size.  For 
wide beams, confusion from point sources is a significant
source of noise, 
White \& Majumdar \cite{WhiMaj04}, Knox et al \cite{KnoHolChu03}.
For small beams, several pixels must be combined to produce the
total cluster signal (see for example Battye \& Weller \cite{BatWel03})
above the $Y_{min}$ threshold, 
and there may be errors inherent to the corresponding cluster finding.
These can be dealt with both in the data acquisition 
(e.g. by having 
more frequencies and an appropriate scanning strategy to help identify
the point sources) and in the analysis; the effects particular to
an experiment will depend strongly on the details of that experiment.
An early example finding clusters in a noisy map was done by
Schulz \& White \cite{SchWhi03}, a more recent start-to-finish analysis 
of N-body simulations, including cluster finding and noise modeling, 
has been done for Planck SZ clusters by Geisbusch, Kneissl \& Hobson 
\cite{GeiKneHob04}. See also Melin, Bartlett \& Delabrouille 
\cite{MelBarDel04}, Pierpaoli et al \cite{Pie04}, Vale \& White 
\cite{ValWhi05} for more on cluster
finding.}
Interferometer experiments will ``resolve out'' some of the power and thus
will effectively have a higher $Y_{min}$.
In addition, false clusters detected due to alignments of low mass SZ
sources will need to be discarded via some sort of follow up.
The end result of this processing for an SZ cluster survey will be
a catalogue of clusters (with angular positions) with SZ decrement above 
some minimum threshold value $Y_{min}$. 

We show in figure 1 a plot of the minimum cluster mass for a given $Y_{min}$ as a function of
redshift, for some representative $Y_{min}$ values 
expected with APEX and SPT.  
As mentioned above, the minimum mass depends on
the mass temperature normalization $T_*^{SZ}$ which is not well
known and will be discussed in our section on modeling uncertainties,
section \S 4.2.  For illustration we have
taken representative values for $T_*^{X-ray}$ from X-ray measurements, 
which we might expect to be close to $T_*^{SZ}$.
For $Y_{min}$ we have taken the APEX's
quoted 10 $\mu$K sensitivity and
multiplied by a factor of 5, which would be a naive 5-$\sigma$ detection.
$f_{gas}$ is taken to be 
$0.10 h^{-3/2} M_{15}^{0.148}/(1 + 0.10 M_{15}^{-0.25})$ (Lin, Mohr
\& Stanford \cite{LinMohSta03}), in practice it also
has a scatter.\footnote{The scatter found by Lin et al \cite{LinMohSta03}
gives
\eq
f_{gas}=0.10(\pm 3\%) h^{-3/2} M_{15}^{0.148 \pm 27\%}/(1 + 0.10(\pm 6\%) 
M_{15}^{-0.25 \pm 28\%}) \; .
\en
}
\begin{figure}{Mass cut}
\label{fig:masscut}
\begin{center}
\plotone{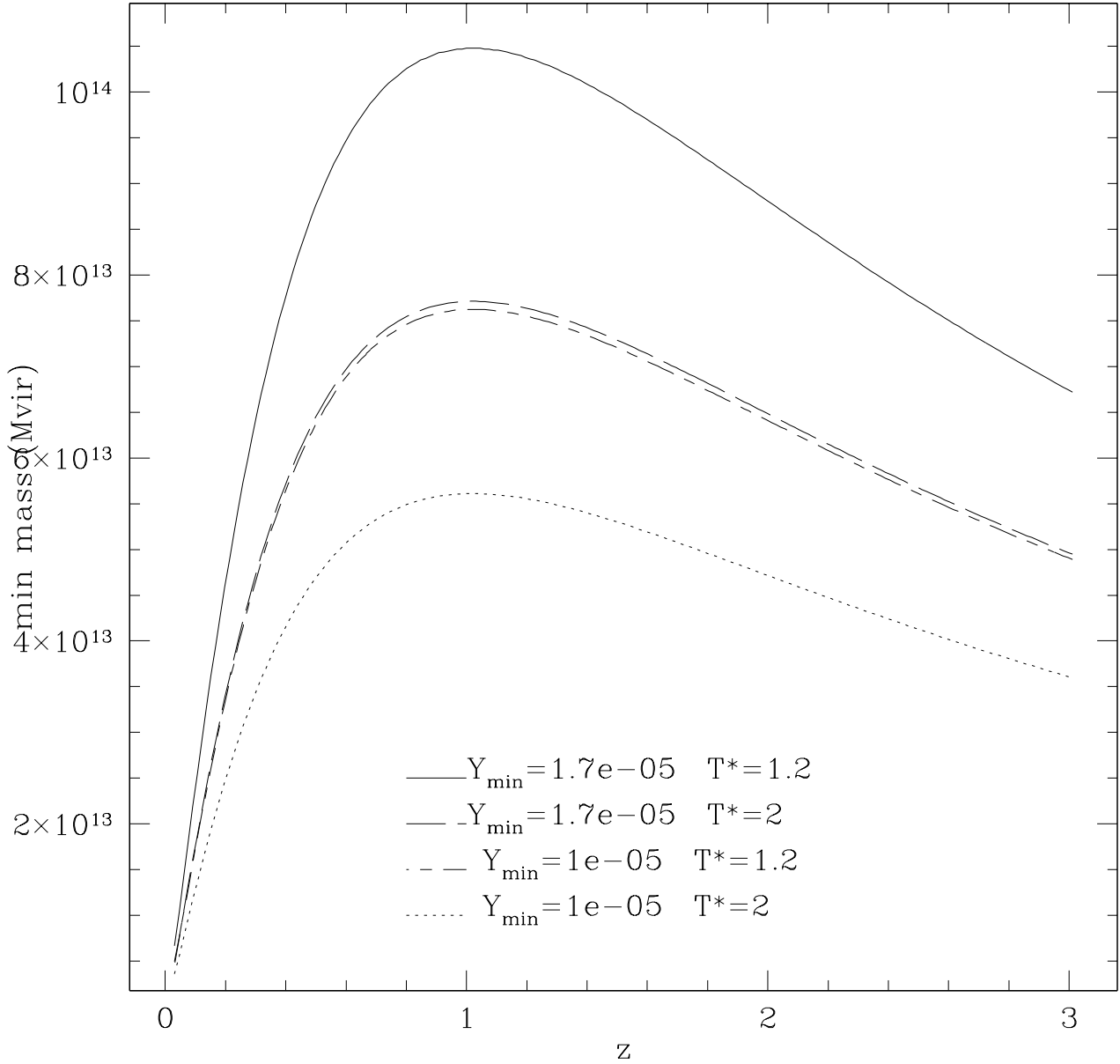}
\end{center}
\caption{Minimum mass for given $Y_{min}$ using equation \ref{ydef}.  Note
that as $Y_{min} \propto T^{SZ}_* (M_{min})^{5/3}$, 
fixed $Y_{min}/T^{SZ}_*$ will give the same $M_{min}$.  At 143 GHz, 
$Y= -\Delta T/T$, at 150 GHz, corresponding to APEX, $Y = -0.96 \Delta T/T$.  
Note that these are virial masses $M_{vir}$, $M_{200}$ will be 
smaller by a factor of 0.77, see the discussion in section \S 4.2 about
mass conversions.}
\end{figure} 
The slow change in $M_{min}$ with redshift is a feature of SZ
selection, which in principle allows clusters of similar masses to 
be observed at all distances
(Bartlett and Silk \cite{BarSil94}, Barbosa et al 
\cite{Baretal96}, Holder et al \cite{Holetal00}, Bartlett \cite{Bar01},
Kneissl et al
\cite{Kne01}, Diaferio et al \cite{Dia03}).  We can easily calculate
observable quantities based on $M_{min}(Y_{lim},z)$.  This gives an
advantage over X-ray where the flux dims rapidly at higher redshifts.
An additional advantage of SZ over X-ray is that the SZ signal strength 
depends upon the density, while the X-ray signal strength depends upon 
the density squared.  Thus X-ray measurements
boost the weight of the cluster core, which has poorly understood 
physics, in the detection.  
Conversely, the detection of SZ based on density means that the SZ 
signal is much more sensitive to line-of-sight
contamination (White, Hernquist \& Springel \cite{WhiHerSpr02}).
Specifically, SZ effects are proportional to
the total (hot) gas mass in the cluster
along the line of sight ($ \Delta T \propto \int n_e T_{gas} dl$).
In theoretical models the
dominant contributions come 
from the region within $\sim 0.2-0.4$ of the cluster virial radius
(Komatsu \& Seljak \cite{KomSel02}).
Note that at low redshifts the SZ selection probes very low mass objects,
where the poorly understood gas physics dominates;
as a result we will take a minimum redshift cut of $z = 0.2$.  Such
a cut could be imposed experimentally in the followup.

\section{Analytic calculations}
In order to go from this cluster catalogue to cosmologically useful
mass counts, theoretical processing and assumptions are needed.
In this section we review and set the notation for the
angular power spectrum/correlation function in terms of analytic
quantities (which will be varied in the next section).
Theoretical inputs to the analytic calculations include the choice of
mass function, transfer function, biasing scheme, mass temperature relations
(in particular $T^{SZ}_*$), the initial power spectrum and of course 
cosmological parameters.  Some of these, such as the mass function,
are well tested, other quantities such as $T^{SZ}_*$, the
mass temperature normalization appropriate for calculating the $Y$ parameter,
are not determined well at all either theoretically or observationally,
and widely varying approximations are in use.  We will compare these
approximations in the following.  
The angular power spectrum/correlation function
can be calculated with analytic prescriptions for
the mass function $dn/dM (z)$ and the bias relating the dark 
matter correlations to the correlations for the galaxy clusters.
The two dimensional correlation function is (Moscardini et al \cite{Mos02},
Diaferio et al \cite{Dia03},Mei \& Bartlett \cite{MeiBar03}):
\eq
w(\theta) =
\frac{\int^{\infty}_0 r_1^2dr_1
\int^{\infty}_0 r_2^2 dr_2 
 \int_{M_{min}(Y_{min})}^\infty dM_1 \int_{M_{min}(Y_{min})}^\infty dM_2
\phi(M_1,r_1)\phi(M_2,r_2)
\xi(M_1,M_2,{\bf r_1 - r_2})}
{[\int^{\infty}_0 r^2 \int_{M_{min}(Y_{min})}^\infty dM\phi(M,r)dr]^2}
\en
where $r_1, r_2$ are the radial distances of the clusters which have
three dimensional positions ${\bf r_1,r_2}$ and ${\bf r_1 \cdot r_2}=
r_1 r_2 \cos \theta$.  
$\xi(M_1,M_2,{\bf r_1 - r_2})$ is the three dimensional cluster
correlation function and $\phi(M,r)$ is the selection function
(as a function of radial distance). 
In particular, 
\eq
\begin{array}{l}
\phi(M,r) = \frac{dn}{dM} (z(r)) \\
\xi(M_1,M_2,{\bf r_1 - r_2}) = 
b(M_1,z_1(r_1)) b(M_2,z_2(r_2)) \xi_{dm}({\bf r_1 - r_2}) \\
\xi_{dm}({\bf r_1 - r_2}) = \xi_{dm}(r) = 
\int \Delta^2(k) \frac{\sin kr}{kr} d \ln k \;, \; r = |{\bf r_1 - r_2}|  
\\
\Delta^2(k) = \frac{V}{(2 \pi)^3} 4  \pi k^3 P(k) \\ 
P_{lin}(k)= D^2(z) P_0 k^n T^2(k)
\end{array}
\en
In the above, 
the selection function is the number of clusters as a function
of redshift (with the normalization included explicitly).
The linear power spectrum $P_{lin}$ comes from an
initial power spectrum with slope $n$, normalization
$P_0$ implied by our choice of $\sigma_8$, and transfer function $T^2(k)$.
The nonlinear power spectrum is derived from $P_{lin}$
using the method of Smith et al \cite{Smi02}.
The linear bias $b$ is also defined above, nonlinear 
bias will be considered in the next section.

The full two-dimensional correlation function thus becomes
\eq
\label{wtheta}
\displaystyle
\begin{array}{ll}
w(\theta) &=
\frac{\int^{\infty}_0 r_1^2dr_1
\int^{\infty}_0 r_2^2 dr_2 
\int_{M_{min}(Y_{min})}^\infty d M_1 \frac{dn}{dM_1} (z(r_1)) 
\int_{M_{min}(Y_{min})}^\infty d M_2 \frac{dn}{dM_2} (z(r_2)) 
b(M_1,z_1(r_1)) b(M_2,z_2(r_2)) \xi_{dm}({|r_1 - r_2|}) }
{[\int^{\infty}_0 r^2\int_{M_{min}(Y_{min})}^\infty d M \frac{dn}{dM} (z(r)) 
dr]^2} \\
&= \int^{\infty}_0 r_1^2\Phi(r_1)
\int^{\infty}_0 r_2^2 dr_2  \Phi(r_2)
\xi_{dm}({|r_1 - r_2|}) 
\\
\end{array}
\en
where we have defined
a generalized selection function (Moscardini et al \cite{Mos02},
Mei \& Bartlett \cite{MeiBar03}\footnote{
Note our selection function differs from that in Mei \& Bartlett 
\cite{MeiBar03} by the factor $r^2 dr/dz $, their expression for
$w(\theta)$ is equivalent.})
\eq
\label{phidef}
\Phi(r) = 
\frac{\int_{M_{min}(Y_{min})}^\infty d M \frac{dn}{dM} (z(r)) 
b(M,z(r))} 
{[\int^{\infty}_0 r^2\int_{M_{min}(Y_{min})}^\infty d M \frac{dn}{dM} (z(r)) 
dr]} 
\en

The above angular correlation function (and corresponding power spectrum)
can be simplified via
the Limber approximation \cite{Lim53}.  As $\Phi$ varies slowly relative
to the correlation function, the integration can be rewritten as
an integration over an average distance $y \equiv (r_1 + r_2)/2$
and one over relative separations $x = (r_1 - r_2)$.  The integration
over $x$ then gives a Bessel function $J_0$: 
\eq
\begin{array}{ll}
w(\theta)
&\sim \int dy y^4 \Phi(y)^2 \int dx \xi_{dm}(\sqrt{y^2 \theta^2 + x^2}) \\
&
= \int dy y^4 \Phi(y)^2 
\int d \ln k  \pi J_0(k y \theta) \frac{\Delta^2(k)}{k}
\end{array}
\en
Then the  power spectrum 
\eq
\label{Cell}
C_\ell = 2 \pi^2 \int dy y^5 \Phi(y)^2 \frac{\Delta^2(\ell/y)}{\ell^3} \;
\en
can be read off in the small angle approximation
\eq
\begin{array}{l}
w(\theta)=\frac{1}{4 \pi}\sum_{\ell}
\sum_{m=-\ell}^{m=+\ell}|a^m_\ell|^2P_\ell(cos\theta)\\
=\sum_{\ell} \frac{2\ell+1}{4 \pi}
C_\ell P_\ell(cos\theta) \\
\simeq \int d \ell \frac{\ell}{2 \pi} C_\ell J_0(\ell \theta)
\end{array}
\en
because $P_\ell (\cos \theta) \simeq J_0 (\ell \theta)$ for small angle.
We can also define the inverse via 
\eq
C_\ell = 2 \pi \int_0^\infty w(\theta) J_0(\ell \theta) \theta d \theta
\en
The correlation function $w(\theta)$ and its power spectrum $C_\ell$ can be 
transformed to each other by the above equation, and therefore they encode the 
same information.  However, to understand possible measurements
and errors, use of $C_\ell$ is usually preferable because the errors for
different $\ell$ values are uncorrelated for small $\ell$.  We'll
use $C_\ell$ for the most part in the following.

\begin{figure}{``Vanilla'' model}
\label{fig:vanilla}
\begin{center}
\plotone{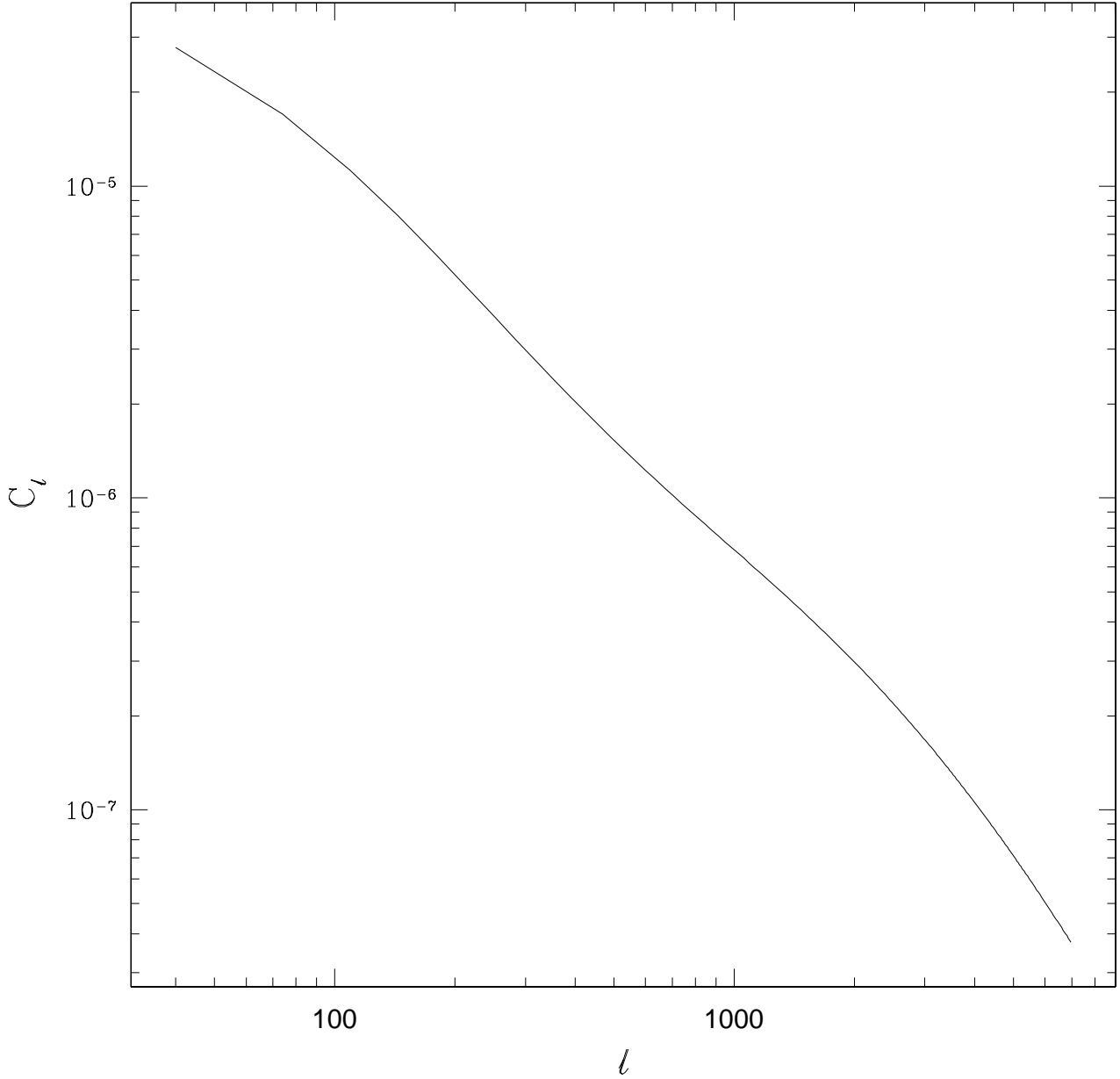}
\end{center}
\caption{Our reference ``vanilla'' model.  The Evrard mass function,
Sheth-Tormen bias, Eisentein-Hu transfer function, 
$T_*^{SZ} = 1.2$, $f_{gas} = 0.10 h^{-3/2}M_{15}^{0.148}/(1 + 0.10 M_{15}^{-0.25})$, 
$\sigma_8 = 0.9, n=1, \Omega_m = 0.3, 
\Omega_b h^2 = 0.02$ and for experimental parameters 
$Y_{min} = 1.7 \times 10^{-5}$ and minimum redshift $z_{min}= 0.2$.
Mass conversions have been taken into account with the prescription
given in White \cite{Whi01}.
}
\end{figure}

For reference, we show in figure 2 the angular power spectrum 
for a representative model.
The dependence on the possible reasonable choices for the analytic and
cosmological model parameters is the subject of a later section.
The choices taken here will be our ``vanilla'' model:\footnote{Note
that this is not identical to the ``vanilla'' model of
Tegmark et al\cite{Tegetal04}, we merely use the term to denote a model 
which we take to be the simplest in some rough sense.  For example, our
``vanilla'' model includes a choice of $T_*^{SZ}$ and of mass function.} 
 for the
dark matter we use
the Evrard mass function, the Sheth-Tormen bias and
the Eisenstein-Hu transfer function, for
cluster parameters we take 
$T_*^{SZ} = 1.2$, $f_{gas} = 0.10 h^{-3/2}M_{15}^{0.148}/(
1+0.10 M_{15}^{-0.25})$, for
cosmological parameters we take $\sigma_8 = 0.9$, $n=1$, $\Omega_m = 0.3$, 
$\Omega_b h^2 = 0.02$ and for experimental parameters we take 
$Y_{min} = 1.7 \times 10^{-5}$.  Our minimum redshift is $z_{min} = 0.2$.
We will use the same axes for (almost) all the plots, so that the
relative impact of different effects are easily comparable.
For the ``vanilla'' parameters one gets about 10 clusters per square degree.

We have made some changes from earlier similar works in this review section.
The basic expression above has been used by Mei and Bartlett \cite{MeiBar03}
and a variant (to be discussed later) has been derived and used 
by Diaferio et al \cite{Dia03}. Figure 2 has the following changes
from this earlier work: it gives the power spectrum rather than the
correlation function, implements mass conversions
(which can change masses by 30\%), and includes
a 10\% scatter in the relation of minimum mass and $Y_{min}$ due to the cluster parameter uncertainties.
Mass conversions and the origin for the amount of scatter we have
chosen are discussed in \S 4.
We also differ Mei and Bartlett in using the Eisenstein-Hu transfer function.

\section{Uncertainties}
\label{uncert}
We now compare the effects of cosmological, modeling, and observational
uncertainties/unknowns on the cluster angular power spectrum.
Some aspects of these, with different assumptions, have
been considered previously for the cluster angular correlation function:
Diaferio et al \cite{Dia03} consider two cosmological models
and vary $Y_{min}$ and the bias, Mei and Bartlett \cite{MeiBar03,MeiBar04}
vary $Y_{min}$, $\sigma_8$, $\Omega_m$ and $T_*^{SZ}$.  For the
spatial power spectrum in a Fisher matrix analysis, 
Wang et al \cite{Wan04}
vary these and the primordial fluctuation spectrum, the dark energy 
density and equation of state, the baryon density.  Both they and
 Mei and Bartlett \cite{MeiBar03,MeiBar04} 
allow extra $(1+z)$ and $M$ factors to appear in the $Y(M)$ relation.
Mei and Bartlett find a small effect, note they are at
(large relative to our work here) values of the $Y$ parameter, corresponding 
to large masses where the gas physics isn't as important.
Majumdar \& Mohr \cite{MajMoh03,MajMoh03a} vary most of these
factors as well in finding their constraints.  Except for the
Mei \& Bartlett \cite{MeiBar04} paper which considers APEX, 
these other experiments
are primarily concerned with far future experiments such as SPT.

In this work we consider other theoretical uncertainties such as
changing the mass function and the gas fraction.  Another
new aspect of this work is comparing all these recognized 
uncertainties to each other, which helps identify which modeling
and observational uncertainties need to be reduced the most.  In addition,
as mentioned earlier, we are primarily concerned with the angular
power spectrum as in the last two works, 
rather than the correlation function.  Unlike the angular power
spectrum in the Gaussian regime,
the errors in the correlation function $w(\theta)$ are correlated which makes 
it more difficult to understand how well measurements at given 
separations can determine various quantities.

\subsection{Cosmological model dependence}
We start by showing what changes to the cosmological model do to
the ``vanilla'' model, for later comparison with the
modeling and experimental uncertainties.
For instance, the current published joint WMAP/SDSS cosmological
parameters and errors are $\Omega_m = 0.30 \pm 0.04$, 
$\sigma_8 = 0.86^{+0.18}_{-0.11}$ (Tegmark et al \cite{Tegetal04}). 
We compare our ``vanilla'' model with changes in $\Omega_m$ and $\sigma_8$,
in figure 3.
\begin{figure}{Cosmological parameter dependence}
\label{fig:cosmol}
\begin{center}
\plotone{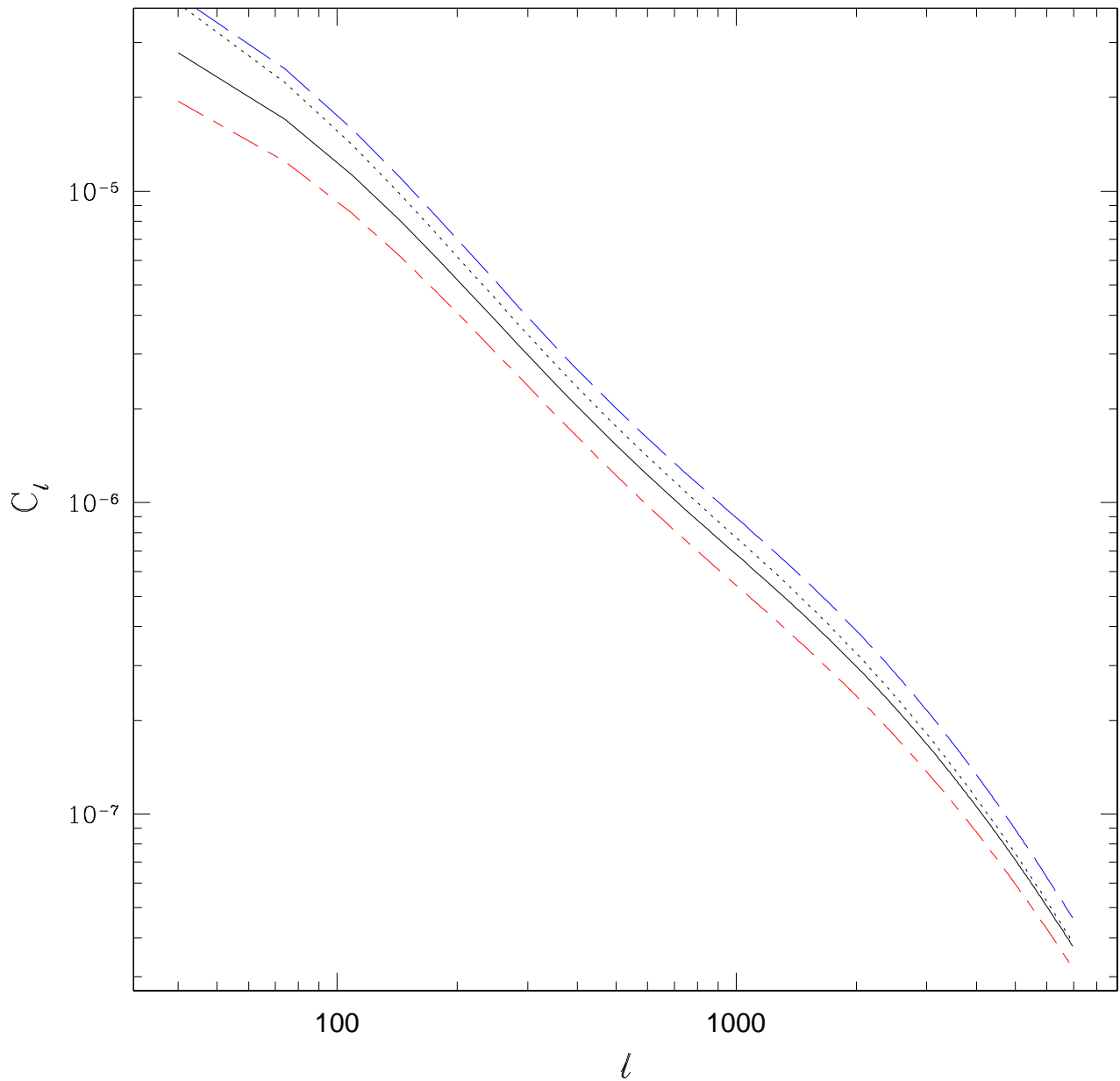}
\end{center}
\caption{The dependence of the power spectrum on varying $\sigma_8$ and
$\Omega_m$ separately.  The solid line is our reference vanilla model
with $\sigma_8 = 0.9, \Omega_m = 0.3$.
The long dashed line is for $\sigma_8= 0.8$, 
the short-long dashed line is for
$\sigma_8 = 1.0$ and the dotted line is for
$\Omega_m = 0.25$.
}
\end{figure} 
The correlation function decreases with increasing $\sigma_8$, as a
higher $\sigma_8$ means the generalized selection function broadens and moves
its peak to higher redshift\footnote{Mei \& Bartlett \cite{MeiBar03} 
have illustrated this effect in their paper.}.  The broadening gives
more non-correlated clusters nearby any given cluster, in addition
the biasing is weaker for high $\sigma_8$.\footnote{We thank the
referee for emphasizing the second point.}
The cluster power spectrum scales quite differently with $\sigma_8$
than the temperature correlation function.  If one parameterizes $C_l$ as 
\ba
C_l(\sigma_8)=\sigma_8^{-\alpha(l)} f(l),
\ea
for $Y_{min}= 1.7 \times 10^{-5}$ and $\sigma_8=0.9$, 
$\alpha(\ell)$ ranges from around 2.7 for $\ell \sim 100$ to
about 1.6 for $\ell \sim 6000$. Doubling $Y_{min}$ changes this
range to 3.5 and 1.7 respectively 
(note the negative power of $\sigma_8$ which is different from the positive 
power scaling behavior of 
the SZ temperature power spectrum (Komatsu \& Seljak \cite{KomSel01}, 
Sadeh \& Raphaeli \cite{SadRep04}).

In fact, the effect of changing $\sigma_8$ is somewhat smaller, 
as there are other constraints which must be satisfied when
changing $\sigma_8$, for instance the observed and hence
fixed number counts of clusters.
One can include this constraint by requiring the number of clusters
to remain fixed along with $\Omega_m$ for instance, and find the
required scaling between $T^{SZ}_*$ and $\sigma_8$(Sadeh \&
Rephaeli \cite{SadRep04}).  Allowing $\Omega_m$ to vary still
gives a constraint using identical
arguments to Huterer \& White \cite{HutWhi02} (also see
Evrard et al \cite{Evretal02}) who were considering
the X-ray temperature:  Fixing the number of clusters with 
a given $Y$ parameter (an observable) and allowing
$T^{SZ}_*, \sigma_8$ and $\Omega_m$ to vary gives
the following (directly analogous to X-ray) 
scaling relation,
\eq
T^{SZ}_* \sim (\sigma_8 \Omega_m^{0.6} )^{-1.1} \; .
\en
More precisely it will be
$f_{gas} T^{SZ}_*$ on the left.  The modeling parameter $T_*^{SZ}$ will
be discussed in detail in the next section.
This relation already suggests a strong degeneracy between the
effects of changing $\sigma_8, T_*^{SZ}, \Omega_m$, see section
\S 4.3 for more discussion on degeneracies.

\subsection{Modeling Uncertainties}
There are several parameters and functions that go into the theoretical
predictions besides the cosmological parameters.  These can be divided
into those related to cluster properties independent of the SZ effect,
and those related to the transformation from cluster mass to the 
observable $Y$ (equation \ref{ydef}).
We will treat both of these in turn.
For an SZ selected survey, uncertainties in modeling affect results
by bringing objects into or out of the survey.
As a result, uncertainties that have a large effect on the observational
properties of the lowest mass clusters in the survey 
($M_{vir} \sim 10^{14} h^{-1} M_\odot$) have the most impact.

{\bf Dark matter cluster properties:}
The cluster properties independent of the SZ effect used in
the analytic prediction of $C_l$, equation \ref{Cell}, are
the correlation function of the dark matter, the mass function, and the bias.

{\it Dark Matter correlation function:}
The dark matter correlation function is quite well known, and can be
obtained from the initial power spectrum via a transfer function
(such as BBKS \cite{BBKS} or Eisenstein \& Hu \cite{EisHu97}) and
then by implementing a nonlinear power spectrum prescription 
(such as Peacock and Dodds \cite{PD96} or Smith et al \cite{Smi02}).
The vanilla model uses the more accurate recent fits, i.e., the nonlinear 
power spectrum fit by Smith et al and
the Eisenstein \& Hu transfer function (the latter is within 
3\%  of the exact CMB power spectrum, M. White, private communication). 
We compare the case with the BBKS transfer function to the vanilla model, 
as it is still in use by many groups.

{\it Mass conversions:}
Before defining the mass functions, it should be recalled that there are 
several different mass definitions in use.  For SZ calculations, the 
mass-temperature relation usually involves the virial mass, but the 
popular mass functions usually are instead for some linking length 
or parameter
$\Delta$ such that the mass inside a radius $r_\Delta$ is
$\Delta$ times the critical density\footnote{Note that some
people use $\Delta$ to refer to the density relative to the
mean density.}:
\eq
\label{mdelta}
M_\Delta = \frac{4 \pi}{3} \Delta \rho_{crit} r_\Delta^3 =
\int_0^{r_{\Delta}} 4 \pi r^2 dr \rho(r) \; .
\en
This mass can differ significantly from the virial
mass (White \cite{Whi01}), which enters into the definition
of $Y$, and conversion must be made between the virial mass and the
mass appearing in the mass function.  The way suggested in
White \cite{Whi01} is to start by assuming
an NFW (Navarro, Frenk and White \cite{NFW}) mass profile, a fitting
formula for this method is given in Hu \& Kravtsov \cite{HuKra03}.
For instance, in an $\Omega_m = 0.3$ universe, $M_{vir}/M_{200} \sim 1.3$
for a cluster with concentration 5.  A 30\% difference in mass is 
significant when one is attempt to make percent-level predictions!
Masses enter the calculations in the $Y(M)$ relation, the mass function
and the bias; consistent definitions or the relevant conversions are 
needed.
We show in figure 5 the effect on the vanilla model
from using $M_{200}$ rather than the appropriate masses
in the $Y(M)$ relation and bias.

{\it Mass Functions:}
We compare the effects on the angular power spectrum of the 
above two possibilities, using a less accurate transfer function
and neglecting the mass conversion, and
three commonly used mass functions in figures 4 and 5.

\begin{figure}{Mass functions}
\begin{center}
\plotone{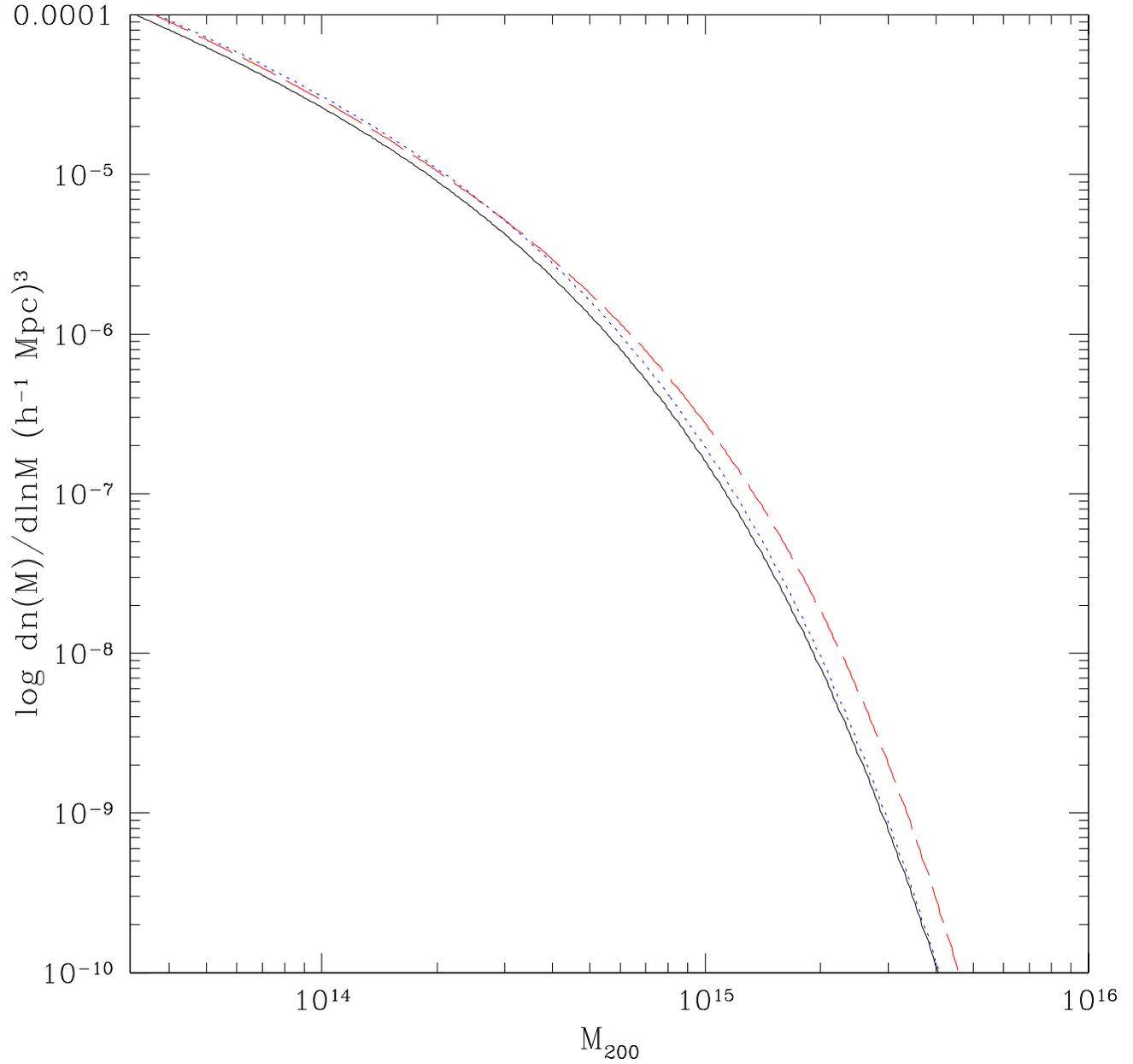}
\end{center}
\caption{ Evrard et al, Jenkins et al and Sheth-Tormen
mass functions.  The smooth line is the Evrard mass function,
the dotted line is the Jenkins mass function and
the dashed line is the Sheth-Tormen mass function.}
\end{figure}
\begin{figure}{Mass function dependence}
\label{compare}
\begin{center}
\plottwo{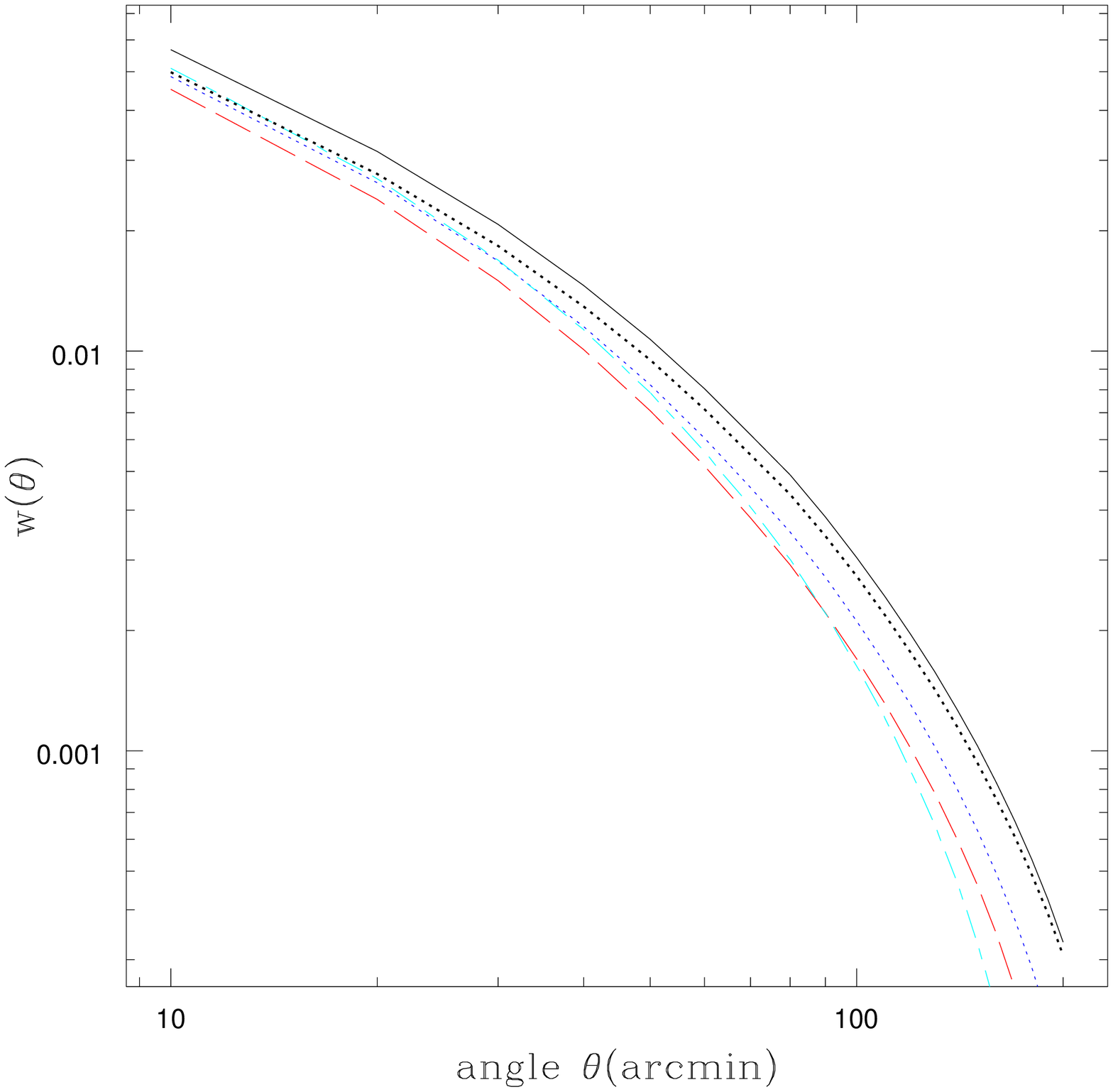}{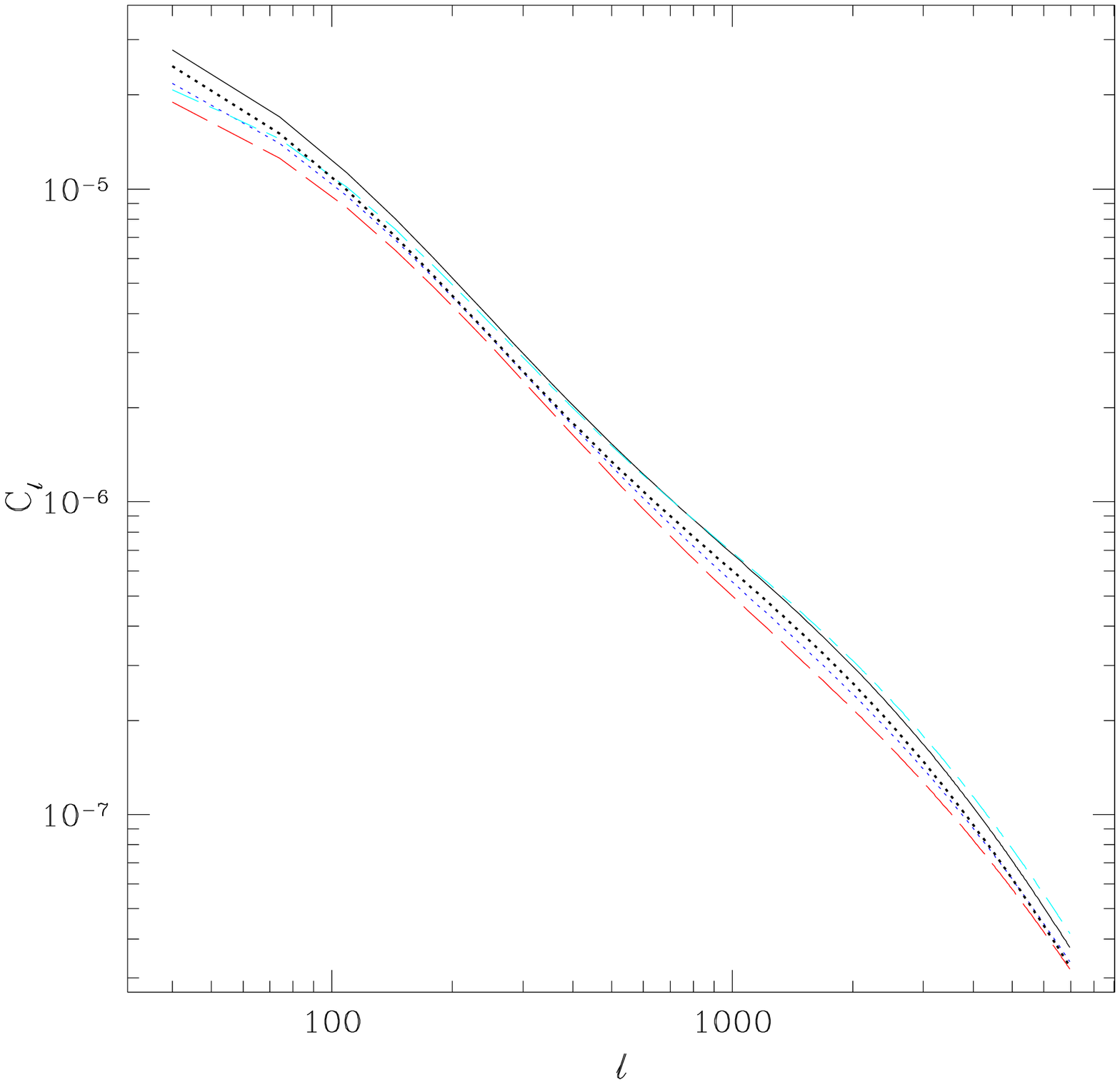}
\end{center}
\caption{On left:
The angular correlation function, same models and line 
labeling as figure 4,  plus
the vanilla model except for BBKS transfer function (dot-dashed line) and 
vanilla model without using mass function conversion for mass-temperature
(heavy dotted line).
On right: the corresponding angular power spectra.}
\end{figure}

The three mass functions taken here can be viewed
as generalizations of the heuristic mass function of
Press and Schechter \cite{PS}.  These are based upon simulations with
large enough volume to accurately sample the number of rare objects such as 
galaxy clusters:
the Jenkins et al \cite{Jenetal01} mass function, the 
Sheth-Tormen \cite{SheTor99} mass function and the Evrard et al
\cite{Evretal02} mass function.  The first two are for masses
in terms of $M_{180 \Omega_m}$, while the last is in terms of $M_{200}$.
These mass functions are expressed in terms of $\sigma (M,z)$ or
$\nu(M,z)$.  Here $\sigma(M,z)$ is the rms of the
mass density field smoothed on a scale 
$R = (3 M/(4 \pi \rho_b))^{1/3}$ with a top hat window function,
$\sigma(M,z) = \int^{\infty}_0 \frac{dk}{k}\Delta^2(k)\tilde{W}_M^2(k,z)
$
where $\tilde{W}_M^2(k,z)$ is the Fourier transform of window function,
$\nu = \delta_c/\sigma$, and $\rho_b = \Omega_m \rho_{crit}$.
We ignore the weak cosmological dependence of $\delta_c$ and 
set $\delta_c \equiv 1.686$.  We then can write
\eq
\frac{dn}{dM} = \frac{\rho_m(z)}{M}
f(\ln \sigma^{-1}) \frac{d \ln \sigma^{-1}}{d M} ;
\en
to get the three mass functions described in Table 1.

\begin{table}[hbtp]
\centering
\caption{}
\begin{tabular}
{|c|c|c|} \hline  
{\rm mass function} & $\Delta \rho_{crit}$& $f(\ln \sigma^{-1}$)\\ \hline
{\rm Sheth-Tormen} & $180 \rho_b$&
$\frac{0.364}{\sigma}(1 + 0.811 \sigma^{0.6})
e^{-\frac{1}{\sigma^2}}$
\\ \hline
{\rm Jenkins} & $180$ $\rho_b$ &$0.301 \exp[-|\ln \sigma^{-1} + 0.64|^{3.82}]$
\\ \hline
{\rm Evrard  et  al}&$ 200$ $\rho_{crit}$ &$0.22 \exp[-|\ln \sigma^{-1} + 0.73|^{3.86}]$\\ \hline
\end{tabular}
\end{table}
The mass density field and the correlation function are derived from the
primordial power spectrum, which we take to be scale free, 
i.e. $n=1$, and normalized by $\sigma_8$.
To get a sense of the effects,
the (less accurate) earlier BBKS \cite{BBKS}
transfer function decreases $C_\ell$ by 5\% at $\ell = 214$,
Neglecting the mass conversion between
$M_{200}$ and $M_{vir}$ decreases $C_{214}$ by 11\% and
the Sheth-Tormen mass function \cite{SheTor99} decreases it by 19\%.
Using the Peacock \& Dodds \cite{PD96} nonlinear prescription 
gives no noticeable change, hence we did not show it in the figure.  
An extensive comparison of different cases is given in the table in section
\S 4.3.

{\it Bias:}
There are also several different possibilities for bias.
The (linear) bias $b(M,z)$ is defined via
\eq
\xi(M,M,r,z) = b^2(M,z) \xi(r,z)
\en
where $\xi(r,z)$ is the dark matter power spectrum and
$\xi (M,M,r,z)$ is the power spectrum of halos of mass $M$.
The original idea of peak biasing by Kaiser \cite{biasedKaiser}
has been improved upon with fits to simulations.
There is the Sheth-Tormen bias \cite{SheTor99}, fit to simulations
and motivated by a moving wall argument, 
the bias found by Sheth, Mo \& Tormen \cite{SheMoTor01} (SMT),
and the bias more recently found by Seljak and Warren 
\cite{SelWar04}.
The Seljak and Warren bias was found for small masses but
has the best statistics currently available.  It overlaps
closely with the Sheth-Tormen bias where it is valid but is
systematically lower and does not
extend very far into the high mass range needed for clusters.
Thus, 
using a combination of the two biases would result in a bias which
doesn't integrate to one when combined with the Sheth-Tormen mass
function.  Consequently we have taken the Sheth-Tormen bias as our default.


\begin{table}[hbtp]
\centering
\caption{}
\begin{tabular}
{|c|c|c|}\hline
bias function &$ \Delta \rho_{crit}$& {\rm bias}\\ \hline
{\rm Sheth-Tormen} & $180 \rho_b$&$ 1 + \frac{1}{\delta_c}(\nu^{'2}-1) +
\frac{2p}{\delta_c(1 + \nu^{'2p})}$
\\  &&
$\nu' = \sqrt{0.707} \delta_c/\sigma = 1.418/\sigma, \; 
p = 0.3$\\ \hline
 Sheth, Mo \& Tormen &$ 180 \rho_b$ &
$1+\frac{1}{\delta_c}
\left[ \nu'^2+b\nu'^{2(1-c)}-\frac{\nu'^{2c}/\sqrt{a}}{ \nu'^{2c}+
b(1-c)(1-c/2)}\right]$ 
\\ 
&& $a = 0.707, b = 0.5, c = 0.6$ \\ \hline
{\rm Seljak \& Warren }& $200 \rho_{crit}$ 
& $0.53 +0.39 x^{0.42} + \frac{0.08}{40 x +1} + 10^{-4}x^{1.7} $\\
&&
$x =M/M_{nl} \; \; M= M_{nl} \leftrightarrow \nu = \sigma  $ \\ \hline
\end{tabular}
\label{biastypes}
\end{table}

For the SZ selected power spectrum 
one integrates over all masses greater
than some $M_{min}(Y_{min},z)$, so that what one is actually
probing is an integral of the bias, i.e. $\Phi(Y_{min},z)$ in equation 
\ref{phidef}.  One can define a related (rescaled by the number
density) quantity:
\eq
b_{method,eff}=\frac{\int_{M_{min}(Y_{min})}^\infty d M \frac{dn}{dM} (z(r)) 
b_{method}(M,z(r))}{\frac{dn}{dm}(z(r))} 
\en
where
$b_{method}$ is one of the above biases.
The linear biasing prescription above doesn't work as well for short 
distances, and for this a ``scale dependent bias''
has been calculated for the cluster correlation functions by
Hamana et al \cite{Hametal01} and Diaferio et al \cite{Dia03}.
This scale dependent bias is also a function of the separation $r$ of
the objects of interest and for Diaferio et al is
\eq
\label{nlbias}
b_{eff}(r,z, Y_{min}) = 
b_{ST,eff}(z, Y_{min})(1 + b_{ST,eff}(z, Y_{min}) \sigma(r,z))^{0.35}
\en
the corresponding expression for Hamana et al has an exponent 0.15.
Diaferio et al have shown that this bias works well for cluster correlation
functions for a range of redshifts.  We will use the Diaferio et al
case for illustration.  The bias of Hamana et al is midway
between the linear biasing case and the Diaferio et al case \cite{Dia03}.
With the Diaferio et al bias, the correlation function becomes
\eq
\label{wthetab}
\begin{array}{ll}
w(\theta) &=
\int^{\infty}_0 r_1^2\Phi(r_1)
\int^{\infty}_0 r_2^2 dr_2  \Phi(r_2)
\tilde{\xi}_{dm}({|r_1 - r_2|}) 
\\
\end{array}
\en
where 
\eq
\tilde{\xi}_{dm}({|r_1 - r_2|}) =
(1 + b_{ST,eff}(z, Y_{min}) \sigma(r,z))^{0.70} 
\xi_{dm}({|r_1 - r_2|}) 
\en
In the Limber approximation one then finds
\eq
w(\theta)
= \int dy y^4 \Phi(y)^2 \int dx 
(1 + b_{ST,eff}(z, Y_{min}) \sigma(\sqrt{y^2 \theta^2 + x^2},z))^{0.70} 
\xi_{dm}(\sqrt{y^2 \theta^2 + x^2}) 
\en
As this doesn't easily allow a rewriting in terms of $J_0(\ell \theta)$,
obtaining the power spectrum $C_\ell$'s is somewhat more difficult.
In figure 6 we show the correlation function $w(\theta)$ at
left and the power spectrum $C_\ell$ at right for the vanilla model and
then the Sheth, Mo \& Tormen linear bias and the Diaferio et al nonlinear
bias.  (Although the Hamana et al bias is also between the
Diaferio et al bias and the vanilla bias, it is not what is shown in
this figure.)  We transformed the difference of $w(\theta)$'s between the 
vanilla model and Diaferio et al model to get the $C_\ell$'s for
the former.  We use the vanilla model parameters for the Diaferio et al
bias (e.g. $T_*^{SZ} = 1.2$ rather than $T_*^{SZ} = 2.0$ as they did).
\begin{figure}{Bias dependence} 
\label{fig:bias}
\begin{center}
\plottwo{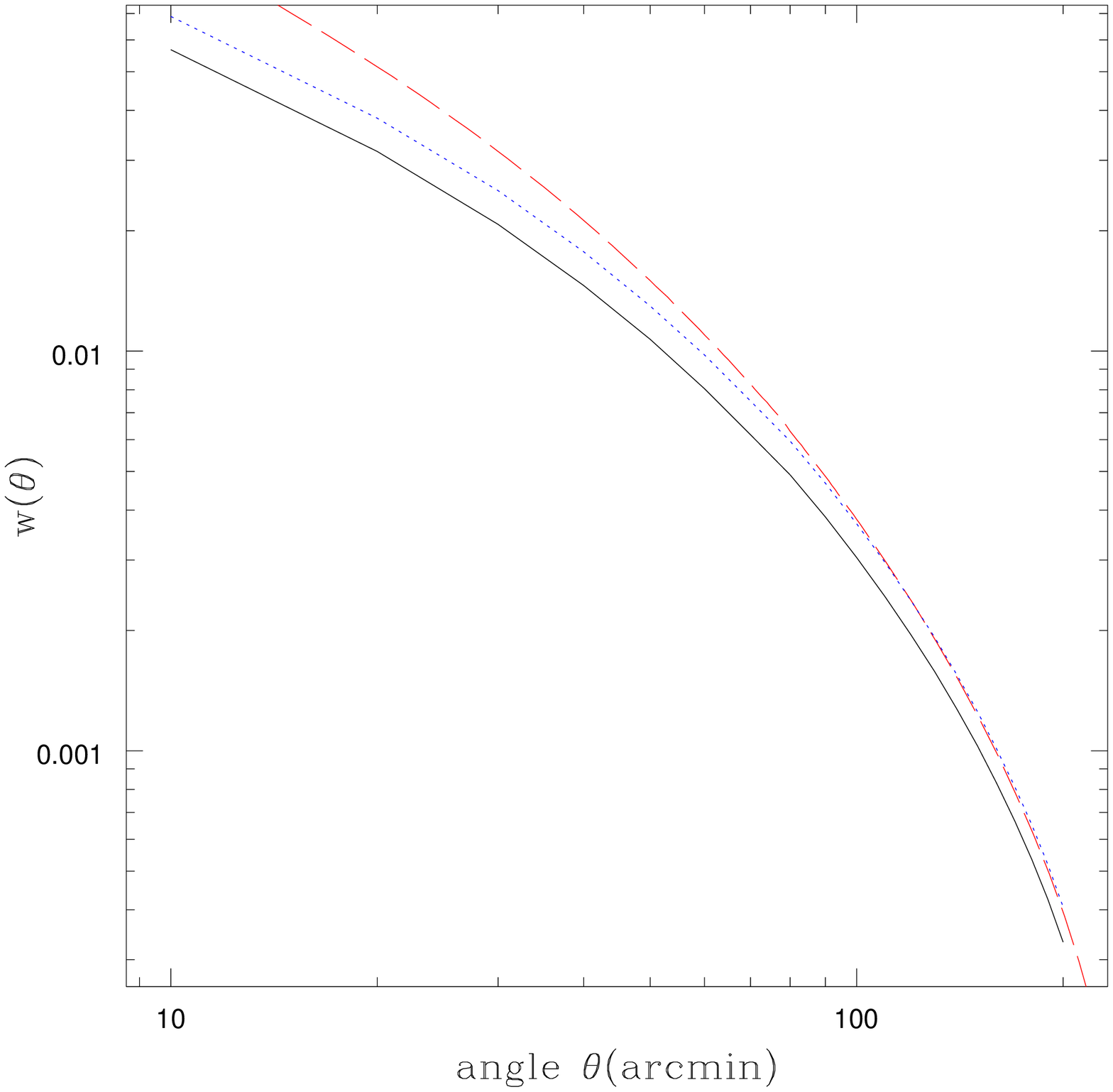}{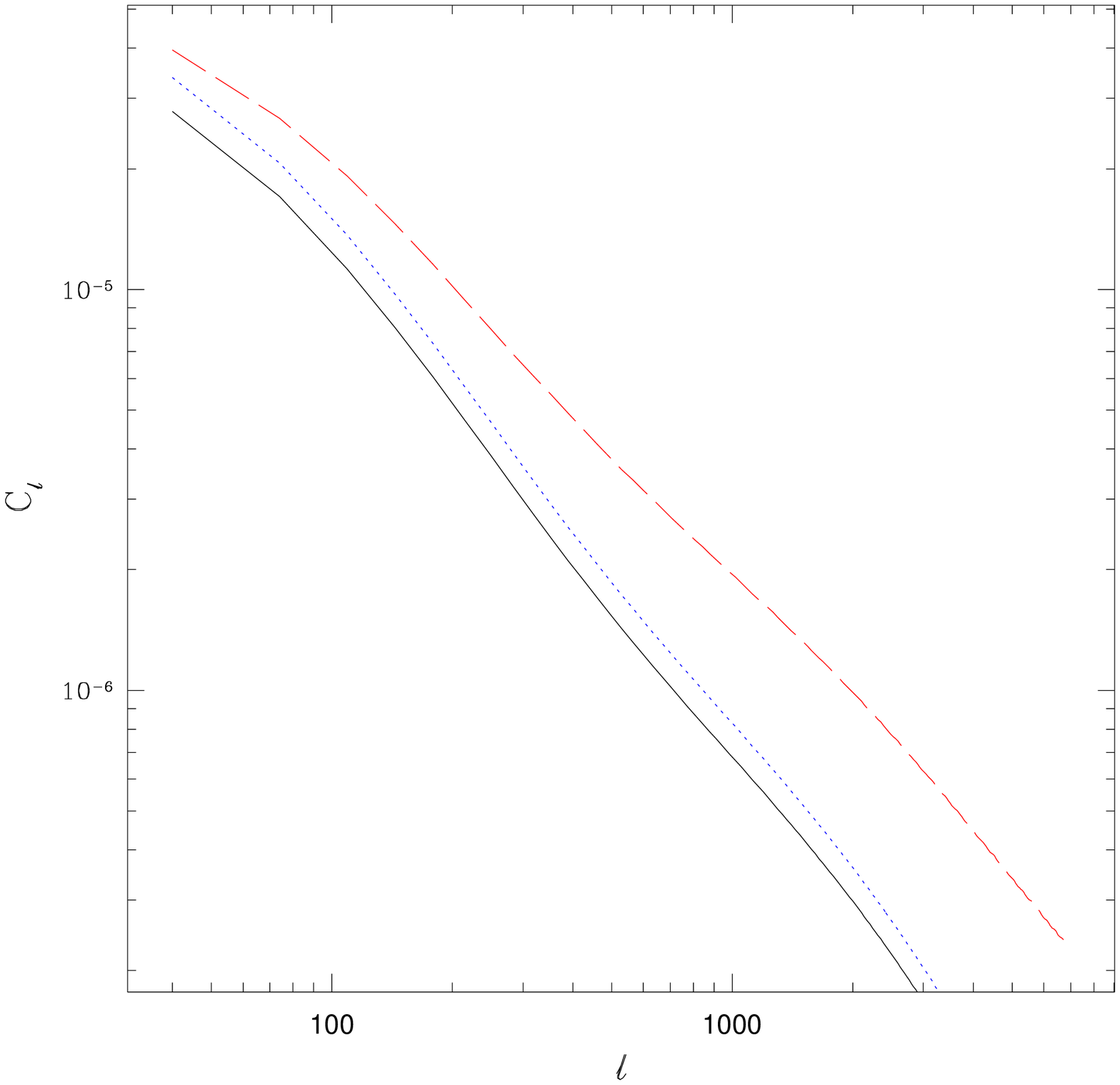}
\end{center}
\caption{The effect of different bias prescriptions on the
correlation function $w(\theta)$ (left) and the power spectrum $C_\ell$
(right).
The vanilla reference model is shown in both cases (solid line), as well as
the cases with the Sheth Mo \& Tormen bias in equation \ref{biastypes} (dotted
line) and
the nonlinear bias of Diaferio et al, equation \ref{nlbias} (dashed line).
}
\end{figure} 
The nonlinear bias has the strongest effect at short distances in the
correlation function--its effect is strongly localized around $\theta
\sim 0$.  Therefore its Fourier transform, the angular power spectrum,
has additional contributions of almost constant magnitude at all $\ell$
relative to the vanilla model.  (The limit of this 
would be adding power at only $\theta =0$, which translates into adding a
constant to the power spectrum.)  Some of this difference is deceptive
as the errors for $C_\ell$ are independent in the linear regime, while
those for $w(\theta)$ combine the $w(\theta)$ measurements for different
values of $\theta$. For a visual comparison, the independence of the
errors for $C_\ell$ makes it easier to draw conclusions about
relations between different parameter choices and uncertainties.
Also, even though the power spectrum and the correlation function are Fourier
transform pairs, estimators for these differ in practice when real data 
is in hand, and give different information when one does not have full 
$2 \pi$ angular coverage in $\theta$.  Ideally one would use both given 
real data.

There is also intrinsic scatter around the bias and the mass functions.
We will not put this in explicitly, however the weak dependence on
the scatter in the $Y(M)$ relation (mentioned below) 
leads us to suspect it will not be a large effect.

{\bf $Y$ parameter:}
The next step is relating the cluster mass to a $Y$ parameter, which
involves more complicated gas physics.  
In addition, the increased sensitivities
of upcoming experiments will allow smaller and smaller mass clusters
to be detected, which are more and more easily disrupted by
this gas physics.  

There are actually three questions: the actual form of the
mass temperature/$Y$ parameter relation, the normalization
of this relation (i.e. $T^{SZ}_* f_{gas}$) and the scatter around this
normalization for a representative group of galaxy clusters.
Simulations alone cannot determine these:
the heating and cooling properties of clusters 
are not understood at an accuracy needed for precision cosmology
and so these questions are intermingled by assumptions used.  For
instance, the scaling relation
was obtained by assuming an isothermal gas profile.  Assuming
hydrostatic equilibrium and using the total $Y$ parameter means
that the details of the profile (many others have been suggested, e.g.
Komatsu \& Seljak \cite{KomSel01}, Loken et al \cite{Lok02}) 
get absorbed into the mass temperature
(or $Y$ parameter) normalization or form.
If the parameters of the gas profile
change with cosmology, it's possible that the normalization will also
do so, or rather the form of the $Y(M)$ relation, a possibility which 
will need to be checked for carefully in the data.
We consider the form of the mass temperature/$Y$ parameter relation,
the normalization, and the scatter in turn.

{\it Mass temperature/$Y$ parameter relation:}
There are several different
mass temperature relations in the literature (see Sadeh \&
Rephaeli \cite{SadRep04} for a description of five common ones),
usually based on X-ray mass temperature relations.  These relations have
been tested both with observations and simulations (bear in mind again
that the simulations do not seem to yet have all the necessary physics),
and some generalizations of these relations have also been tested.
For instance, equation \ref{ydef} can be generalized
to include different $z$ and $M$ dependence, 
such as multiplying by a factor $(1+z)^\gamma M^\alpha$.  
The mass-Y parameter relation also has dependence on $f_{gas}$ 
which can be generalized to change $f_{gas}$ with redshift, or change
it differently with mass.  Observational data and simulation data have
been used to search for these effects.

Most observational tests are of the X-ray mass temperature relations
and of the change of $f_{gas}$ with mass or redshift, rather than
of the $Y(M)$ relation.  For example,  
Ettori et al \cite{Ett04a} have found no evidence for 
additional evolution in redshift of the mass temperature relation.  
For scaling of mass with temperature, Ettori et al \cite{Ett04b} and
Ota \& Mitsuda \cite{OtaMit04} find $T^{X-ray}_{gas} \sim M^{2/3} -
M^{3/5}$, but the departure from the $M^{2/3}$ relation is about
1.5$\sigma$ for the Ettori et al data and
is marginal given the error bars for the Ota \& Mitsuda data.
The former group also
finds marginal (less than $2 \sigma$) evidence for clusters
of a fixed temperature to have smaller gas mass at high redshift.
We have already included the $f_{gas}$ dependence with mass
found by Lin et al \cite{LinMohSta03}, but have not included any
redshift dependence.

More simulations than observations have
addressed the  $z$ and $M$ dependences of the 
$Y(M)$ relation directly.  For example da Silva et al 
\cite{daS04} find numerically that 
the $z$ dependence seems to be well represented by 
the simple scaling given in equation \ref{ydef}.  
For lower mass objects da Silva et al \cite{daS04} find that 
the $M$ dependence in the $Y(M)$ relation steepens from 
$M^{5/3}$.  Our use of an $M$-dependent $f_{gas}$ produces 
an effect in the same direction.  However,
for clusters with $M_{200}$ above $5 \times 10^{13} h^{-1} M_\odot$ 
(or $M_{vir} > 6.5 \times 10^{13} h^{-1} M_\odot$), one
finds that this can just be absorbed into scatter of about 10 \% 
around the $M^{5/3}$ scaling (White et al \cite{WhiHerSpr02}) in
the $Y(M)$ relation. 
The effects of adding an additional factor of
$(1+z)^\gamma M^\beta$ to $Y(M)$ has been considered for the
angular correlation function by 
Moscardini et al \cite{Mos02} 
and Mei \& Bartlett \cite{MeiBar03}, and
Wang et al \cite{Wan04} and Majumdar \& Mohr \cite{MajMoh03, MajMoh03a}
have considered the effects of this additional scaling on parameter
estimation from the
the power spectrum (the three dimensional one in the former case).

{\it Normalization of $Y(M)$ relation:}
For normalization,
we have combined all the mass temperature conversion ignorance into
the parameter combination $Y \propto T^{SZ}_* f_{gas}$.  
The simplest procedure would be to say that 
$Y = Y_{vir}$ (and that gas outside this radius does
not contribute significantly) and to take $T^{SZ}_*$ to be the X-ray value,
\eq
T_*^{SZ} \equiv T_*^{X-ray} \; .
\en
As noted in section \S 2, there is strong disagreement between
simulations and observations for $T_*^{X-ray}$.
Thus the parameter $T_*^{X-ray}$ is not well determined.  Even if it were,
using it to normalize the $Y(M)$ relation is not
necessarily justified.  Not only do X-ray measurements weight more strongly
the center of the cluster, as mentioned before, but for the SZ 
effect the normalization has an
additional contribution due to line of sight contamination from gas 
outside the cluster.  If one takes simulation results (which should be taken 
with a grain of salt given the above mentioned discrepancy), this projection
effect on $Y(M)$ raises the normalization about 8\%
(White et al \cite{WhiHerSpr02}) above the
normalization due to the cluster alone.

The differences between power spectra for different normalizations
of the $Y(M)$ relation are shown in figure 7.
We also show that taking $f_{gas}$ fixed at 0.06 $h^{-1}$ (the value for
a cluster with $M_{vir} = 10^{14} h^{-1} M_\odot$) gives
a power spectrum very close to the vanilla model.  And we additionally 
have shown
a model with  $T_*^{SZ} = 2.2$keV which has the 8\% increase from
line of sight projection (from $T_*^{SZ}=2$keV).
\begin{figure}{Y(M) and mass-temperature normalization dependence}
\label{fig:tstar}
\begin{center}
\plotone{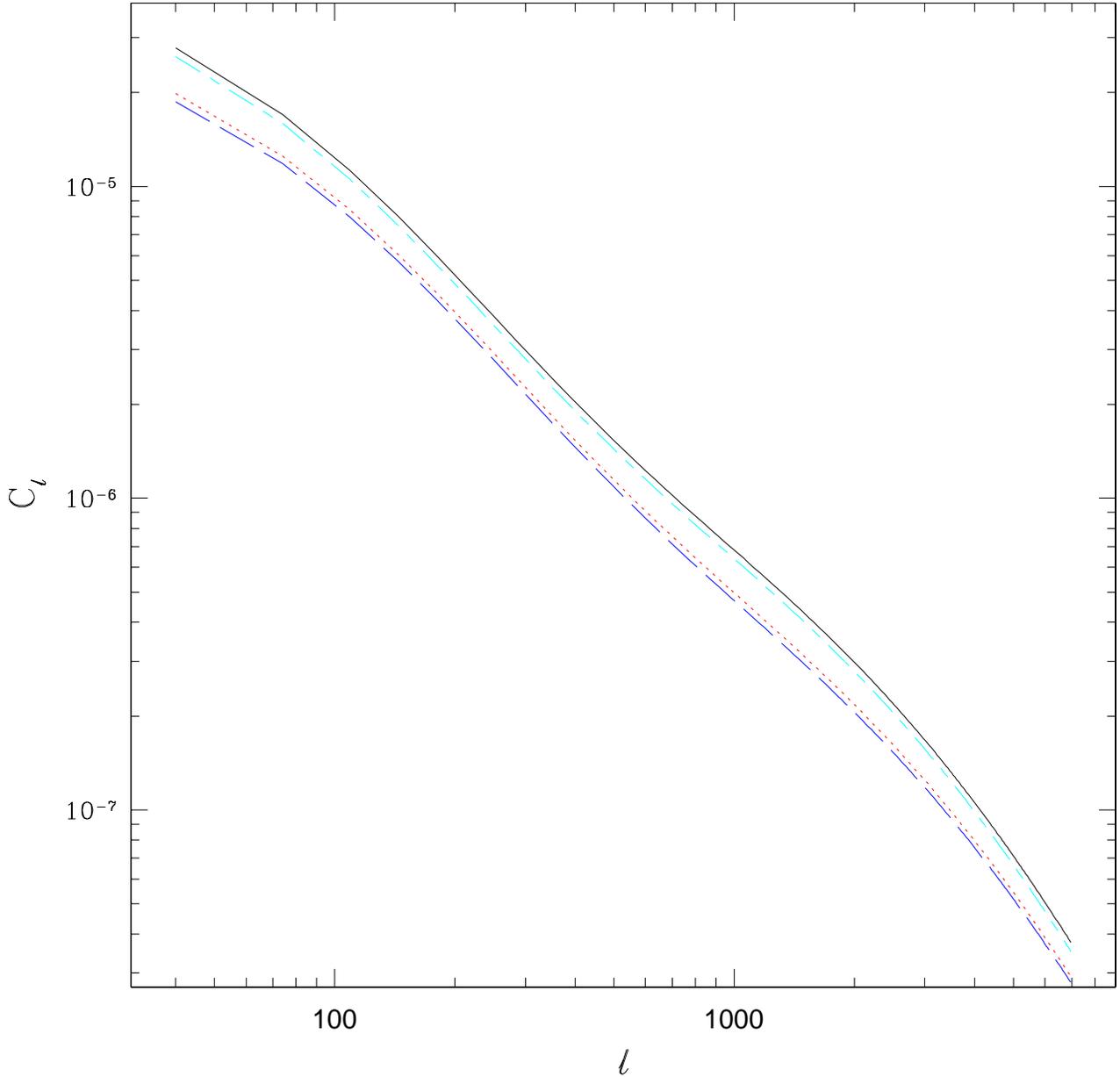}
\end{center}
\caption{The power spectrum for the vanilla model, with 
$T_*^{SZ} = 1.2 $keV (solid line),
and that with $T_*^{SZ} = 2.0$keV (dotted line) as might be suggested by 
X-ray  measurements.  A higher value, $T_*^{SZ} = 2.2$keV (dashed line) is 
also shown to give an idea of the change that an expected SZ line of sight
projection effect (around 8\%) would give.  The dot-dashed line fixes
$T_*^{SZ}=1.2$keV but has no 
evolution in $f_{gas}$ with mass, i.e. $f_{gas} = 0.06
h^{-3/2}$, the value for $M_{vir} = 10^{14} h^{-1} M_\odot$. 
}
\end{figure}
Of course, as changing $T_*^{SZ} f_{gas}$ just rescales the $Y$
parameter, raising $T_*^{SZ} f_{gas}$ is equivalent to lowering
$Y_{min}$, i.e. one is probing clusters with smaller mass.

{\it Scatter:}
Unlike the normalization, the effect of scatter on the $Y(M)$ relation
was very weak.  Taking a 10-12\% 
intrinsic scatter in the mass temperature relation (from
Evrard, Metzler \& Navarro \cite{EvrMetNav96} for X-ray simulations
and similarly from
White et al \cite{WhiHerSpr02} for
SZ simulations) had less than a  percent effect on the $C_\ell$, even with 
APEX sensitivity and accompanying very low mass cuts.
For a large scatter of about 30\%, $C_\ell$ was roughly
decreased by about 3\%.  A similar
robustness to M-T scatter was found in the Fisher matrix calculations
(Levine, Schulz \& White 
\cite{LevSchWhi02}) and in that of number counts 
(Battye \& Weller \cite{BatWel03}).
Metzler \cite{Met98} also found from simulations that the scatter
in the mass temperature relation was larger than that
for the mass-Y parameter relation.
Thus for the bulk of the paper, we have used equation \ref{ydef} and 
combined all our ignorance into the parameters $T^{SZ}_* f_{gas}$.  
We included the unnoticeable 10\% scatter in the Y(M) relation in all our
calculations here.

We were concerned about sources of scatter which are not included
in our analytic description, mergers in particular.
One might expect that processes such as merging will
disrupt the clusters and thus invalidate the assumption of virialization
used in some analytic calculations.  The most massive
clusters have the most recent mergers (as they are generally the
most recently formed objects), but these tend to be included automatically
in the catalogue as their estimated masses, even if inaccurate, are
quite high relative to the mass cut.  The selection for the catalogue 
depends most sensitively on the least massive clusters included, where
mergers are relatively rarer.  (However, at high redshift the
``low mass'' clusters are recently formed as they are the most
massive collapsed objects at that time, so one might expect some
effect from them.)
Mergers are automatically included in the cosmological 
simulations and the scatter in the $Y-M$ relation is not larger
than that expected due to scatter in the $M-T$ relation
(White et al \cite{WhiHerSpr02}), leading us to expect that merger
induced scatter is relatively small.
However, as simulations cannot reproduce cluster properties precisely yet, 
observational data will be needed to calibrate this effect.

\subsection{Degeneracies}

We have considered several different choices for theoretical inputs to the
calculations of the angular power spectrum.  With a brief visual
inspection it can be seen that there are many degeneracies, i.e. that
several changes to parameters and modeling seem to have the same effect on
the power spectrum.\footnote{
We thank the referee for encouraging
us to add more discussion on this issue.}  Note that this does not
take into account other constraints at the same time, such as fixed
number counts, only one parameter is varied at a time.  
We show some of these degeneracies more explicitly in figure 8,
\begin{figure}{Degeneracies}
\label{fig:degen}
\begin{center}
\plottwo{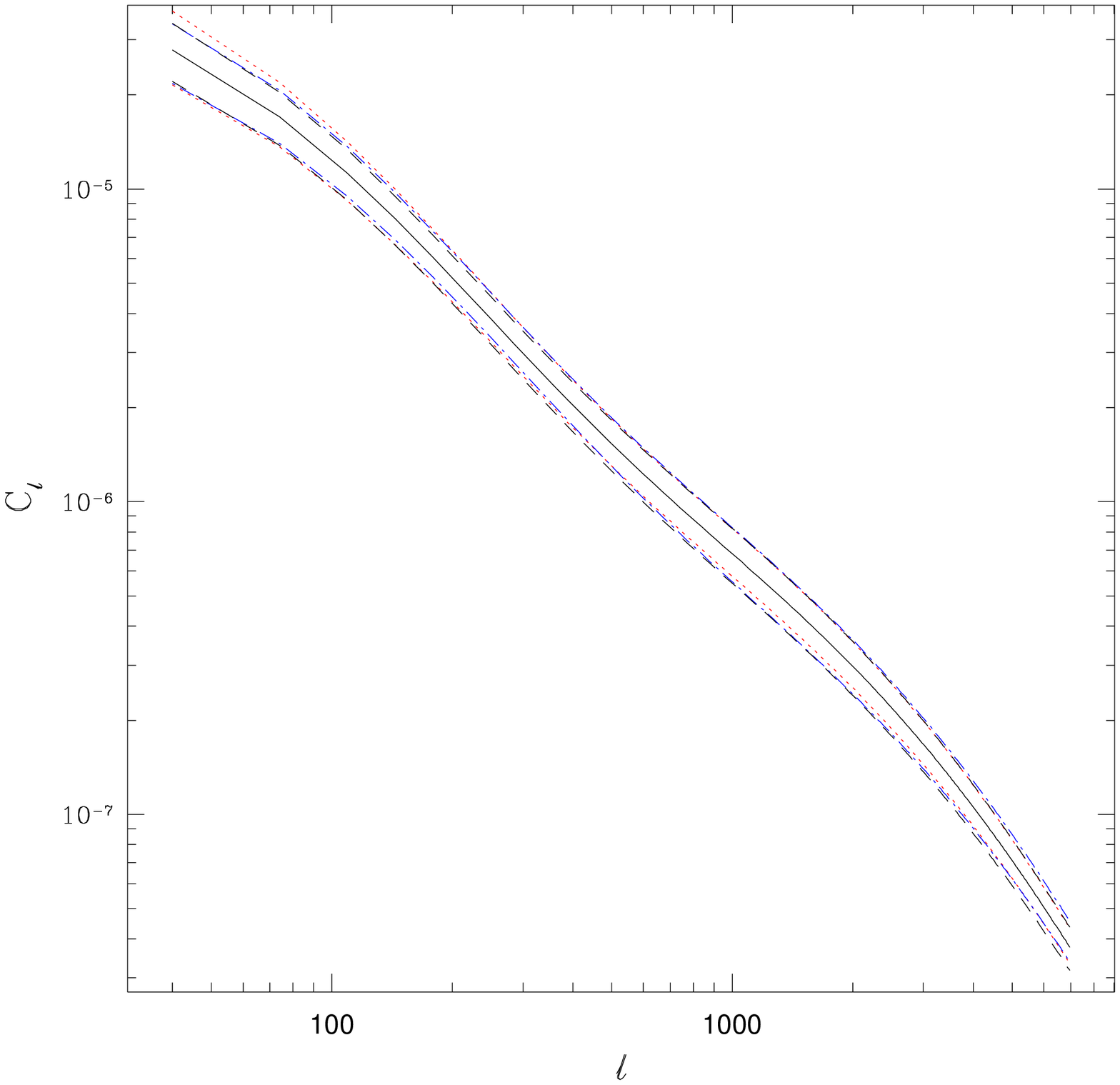}{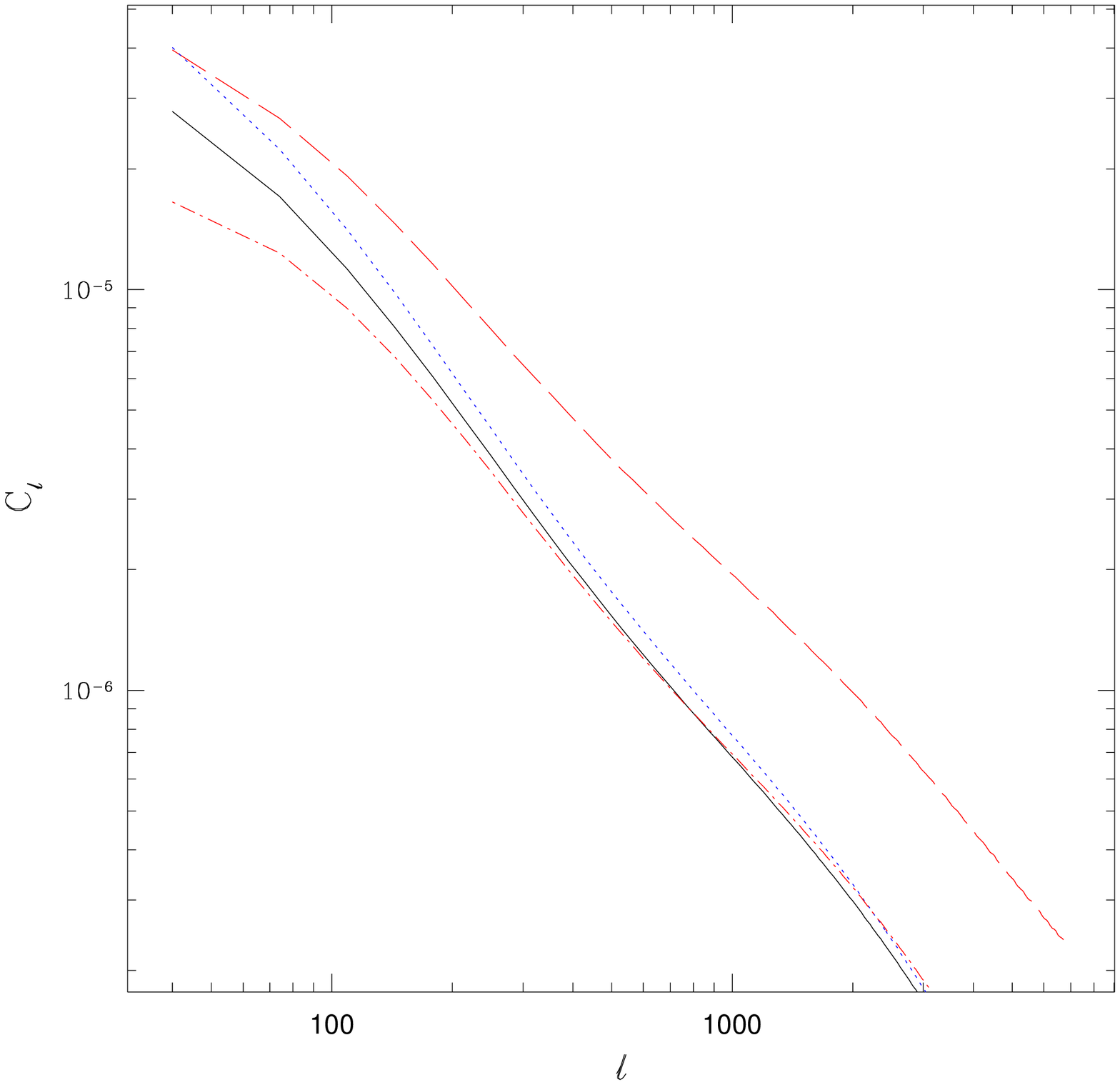}
\end{center}
\caption{Degeneracies in theoretical modeling:
On left, the center line is ``vanilla'' model, the 3 lower lines correspond to
$\sigma_8 = 0.97$ (dotted), Jenkins \cite{Jenetal01} mass function (dot-dash),
and $T_*^{SZ} = 1.7$keV (dashed), and the 3 upper lines correspond to
$\sigma_8 = 0.83$ (dotted), Sheth-Mo-Tormen \cite{SheMoTor01} bias (dot-dash),
and $T_*^{SZ} = 0.9$keV (dashed).
On right, models which are not as degenerate.  The solid
line is ``vanilla'' model, the lower dot-dashed line is the (less accurate)
BBKS transfer function, the upper dotted line is $\Omega_m = 0.25$ and
the top dashed line is the Diaferio et al nonlinear bias.}
\end{figure}
there are other degeneracies in the parameters considered here
that aren't shown.  For instance, choosing a constant $f_{gas}$
rather than having evolution with mass, as done in the previous subsection, is
degenerate with changing $T_*^{SZ}$ from 1.2 keV to 1.3 keV,
using the Sheth-Tormen mass function rather than the Evrard mass
function is degenerate with changing $T_*^{SZ}$ from 1.2 keV to 1.3 keV,
and neglecting the mass conversion from $M_{200}$ to $M_{vir}$ and
$M_{180 \rho_b}$ in the $Y(M)$ relation and bias respectively is
roughly degenerate rescaling $C_\ell$ overall by a factor of 0.88.

This can be compared to changing $\Omega_m$ or using the nonlinear
bias of Diaferio et al, or using a less accurate transfer function
such as BBKS, all of which
change the ``shape'' of $C_\ell$ differently than those above.  These are
shown on the right of figure 8.

Values of $C_\ell$ for these models are compared quantitatively in
the table 3.
\begin{table}[hbtp]
\centering
\caption{}
\begin{tabular}
 {|c|c|c|c|}\hline
{Model} & $ C_{214}/10^{-6}$  &$ C_{1019}/10^{-7}$  & 
$C_{4132}/10^{-7}$\\ \hline
{\rm vanilla}&$4.75$ &$6.67$ &$ 1.00$ \\ 
\hline$ T_* = 0.9 {\rm keV}$&$5.58 $ & $
8.04$ &$ 1.18 $  \\ \hline
$\sigma_8 = 0.83$ &$5.80$  &$ 8.04 $ & 
$1.17$ \\ \hline
SMT bias & $5.75$  &$ 8.10$   &$ 1.21$ \\ \hline

$T_* = 1.7 {\rm keV}$&$ 3.93$& $5.37$& $ 0.82$ \\ \hline 
$\sigma_8 = 0.97$ &$ 3.98$  &$ 5.66$  & $0.87$ \\ \hline
ST mass& 3.87 & 4.89 & 0.79 \\ \hline
Jenkins mass &4.13 &  5.42  & 0.86 \\ \hline
no $f_{gas}$ evol &4.44 &   6.25 & 0. 93 \\ \hline
{\rm no M conversion } &4.16   & 5.89   & 0.88 \\ \hline
{\rm BBKS} & 4.74 & 6.67 & 1.00 \\ \hline
{\rm Diaferio bias} & 9.46 & 1.91&  4.28\\ \hline 
 $\Omega_m = 0.25$ & 5.60  & 7.56 &  1.06 \\ \hline

\end{tabular}
\label{tableofcl}
\end{table}


The fact that $\Omega_m$ is not as degenerate with $\sigma_8$
as expected shows the limitations
of the rough scaling estimate made in section \S 4.1.
Some of these degeneracies can be broken by other measurements,
for instance $dN/dz$, however the degeneracies are very similar,
{\it except for the bias}.  The bias has no effect on $dN/dz$ and thus
can be taken out easily.  In figure 9, we show $dN/dz$ for the same models 
considered for figure 8 (note that we are considering a full steradian).
\begin{figure}{Degeneracies $dN/dz$}
\label{dndzdeg}
\begin{center}
\plotone{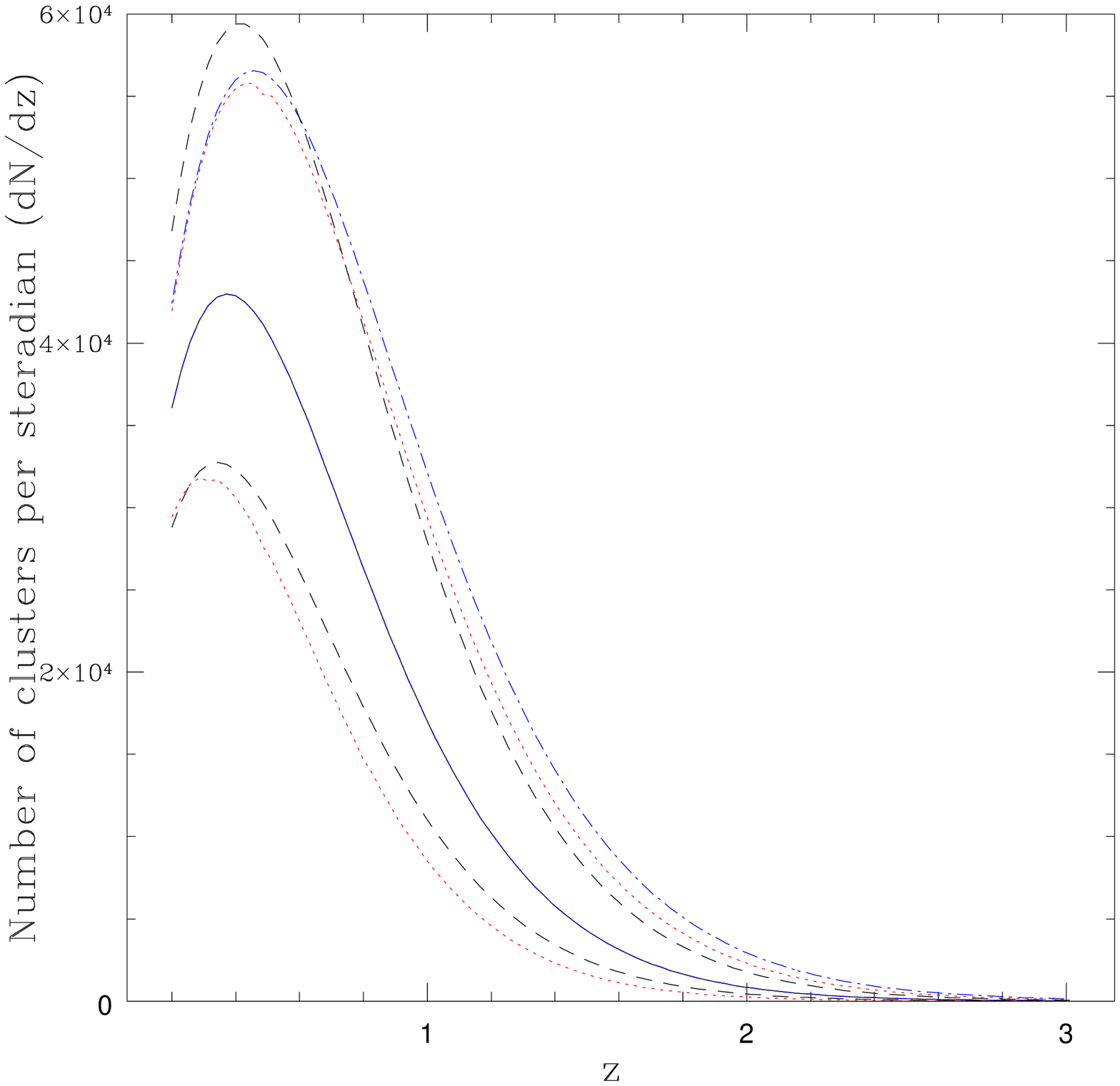}
\end{center}
\caption{Degeneracies in theoretical modeling for dN/dz: labeling
is the same as figure 7 left, except that the upper and lower groups of lines 
have switched.  In addition, changing the bias to that of Sheth-Mo-Tormen
has no effect on $dN/dz$ and so that line is now degenerate with the
vanilla case.}
\end{figure}
As we have shown how choices like the mass function and
different biases are degenerate with other parameter choices (in most
cases), one is only as well measured as the other is known.
In order to see how close these
power spectra are in practice, we now compare to the inherent statistical
measurement error.

\subsection{Cosmic variance, sample variance and shot noise}

There are three sources of inherent statistical measurement error
for the power spectrum $C_\ell$ in the absence of
any systematic errors: shot noise, cosmic variance and sample variance.
These can be combined to give the standard expression for overall
error (see, for example, Knox \cite{Kno95}):\footnote{Here the
shot noise is considered in the Gaussian limit, for the full Poisson
errors for the shot noise, which can be important, see Cohn \cite{Coh05}.
In addition, there are corrections to the error due to the three and
four point functions of galaxy clusters which are usually not included
and we do not include them here.}
\eq
\label{skycoverage}
\delta C_\ell = 
\sqrt{\frac{2}{(2 \ell +1) f_{sky}}} \left(C_\ell + \frac{1}{\bar{n}}\right)  \; .
\en
The factor of $2 \ell +1$ is due to the
$2 \ell +1$ independent measurements of the
power for any $\ell$.  For small $\ell$ (large scales) there are very few 
independent measurements of power in the sky, which is dubbed cosmic 
variance.
Sample variance increases the error as the sky coverage  $f_{sky} \leq 1$
decreases.  Shot noise is determined by $\bar{n}$, the number density 
of clusters per steradian. 

As the depth of the survey goes up ($Y_{min}$ decreasing), the power
spectrum also decreases, as there are more and more clusters of lower
and lower mass, and these are less correlated.
However, the shot noise also goes down.  
On the other hand, as the depth of the survey decreases ($Y_{min}$
increasing), the power spectrum becomes restricted to higher
and higher mass objects and thus goes up.  However as these objects
are rarer, the shot noise also increases.
We can compare the errors
quantities for 3 examples, APEX, SPT, and Planck.  

We use the vanilla model
and again take $Y_{min} = 1.7 \times 10^{-5}$ corresponding
to a 5$\sigma$ detection for APEX with $\delta T \sim 10 \mu$K (at 150 GHz).
APEX will survey 100/200 square degrees at two frequencies (214 and 150 GHz, 
corresponding to 1.4 and 2 mm wavelengths), with 0.75' resolution.
For SPT, we expect about 4000 deg${}^2$ and similar sensitivity and
resolution.  For Planck, we have all sky and will take a rough estimate of
$Y_{min} = 10^{-4}$ (sensitivity/resolution ranging
from  5$\mu$K at 7.0' to 50$\mu$K at 33.0', depending upon frequency).
First we plot the errors due to shot noise,
sampling and cosmic variance for the Planck specifications above
in figure 10.
\begin{figure}{Planck naive errors}
\label{figplanck}
\begin{center}
\plotone{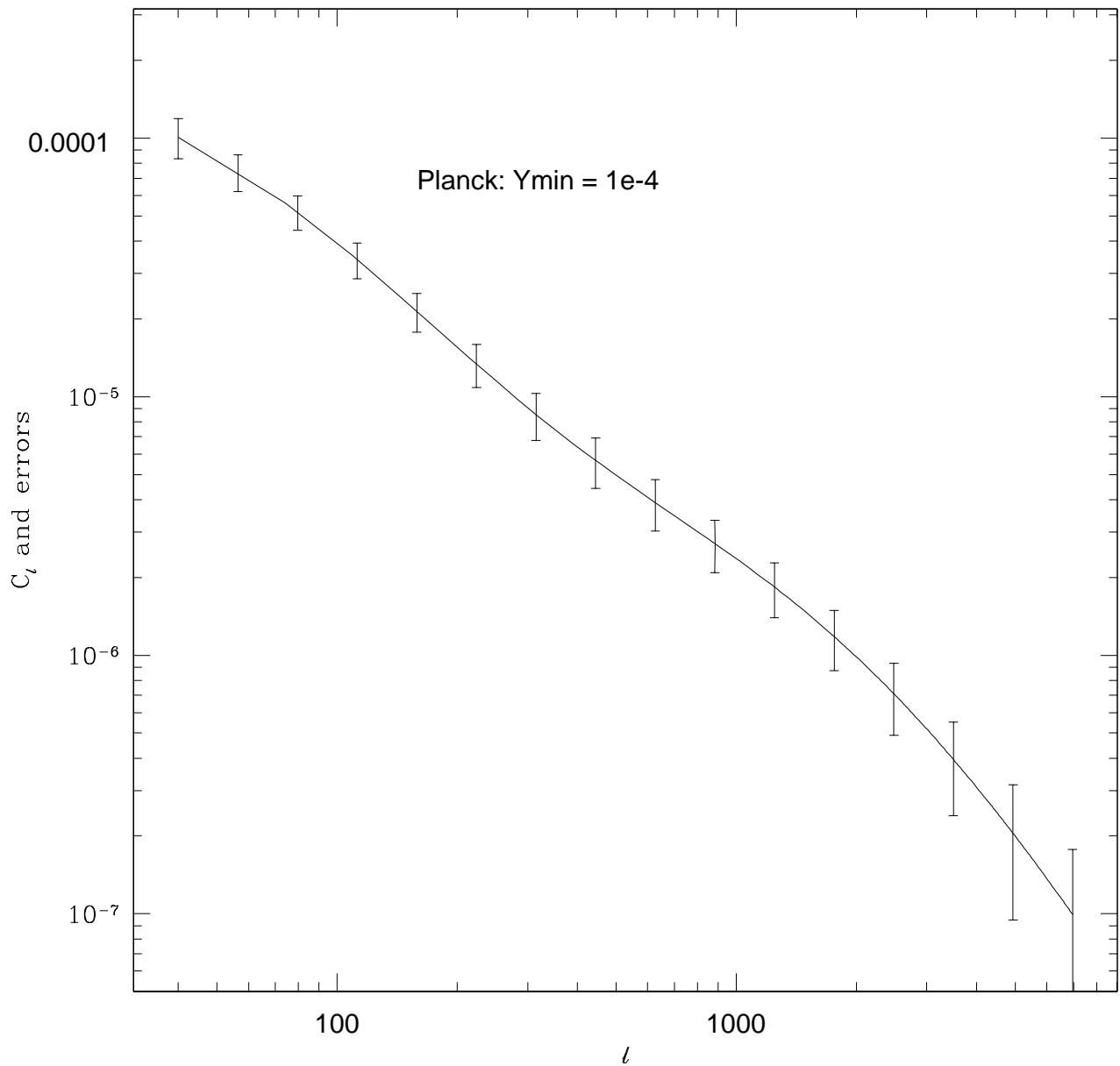}
\end{center}
\caption{The angular power spectrum with errors
from shot noise, sampling and cosmic variance error
expected from Planck, for a $Y_{min} = 10^{-4}$.
The bin size is the spacing between the error bars, equal
spacing in log $\ell$.  See text for more details.
}
\end{figure} 

\begin{figure}{Area vs. sensitivity}
\label{fig:diffarea}
\begin{center}
\plotone{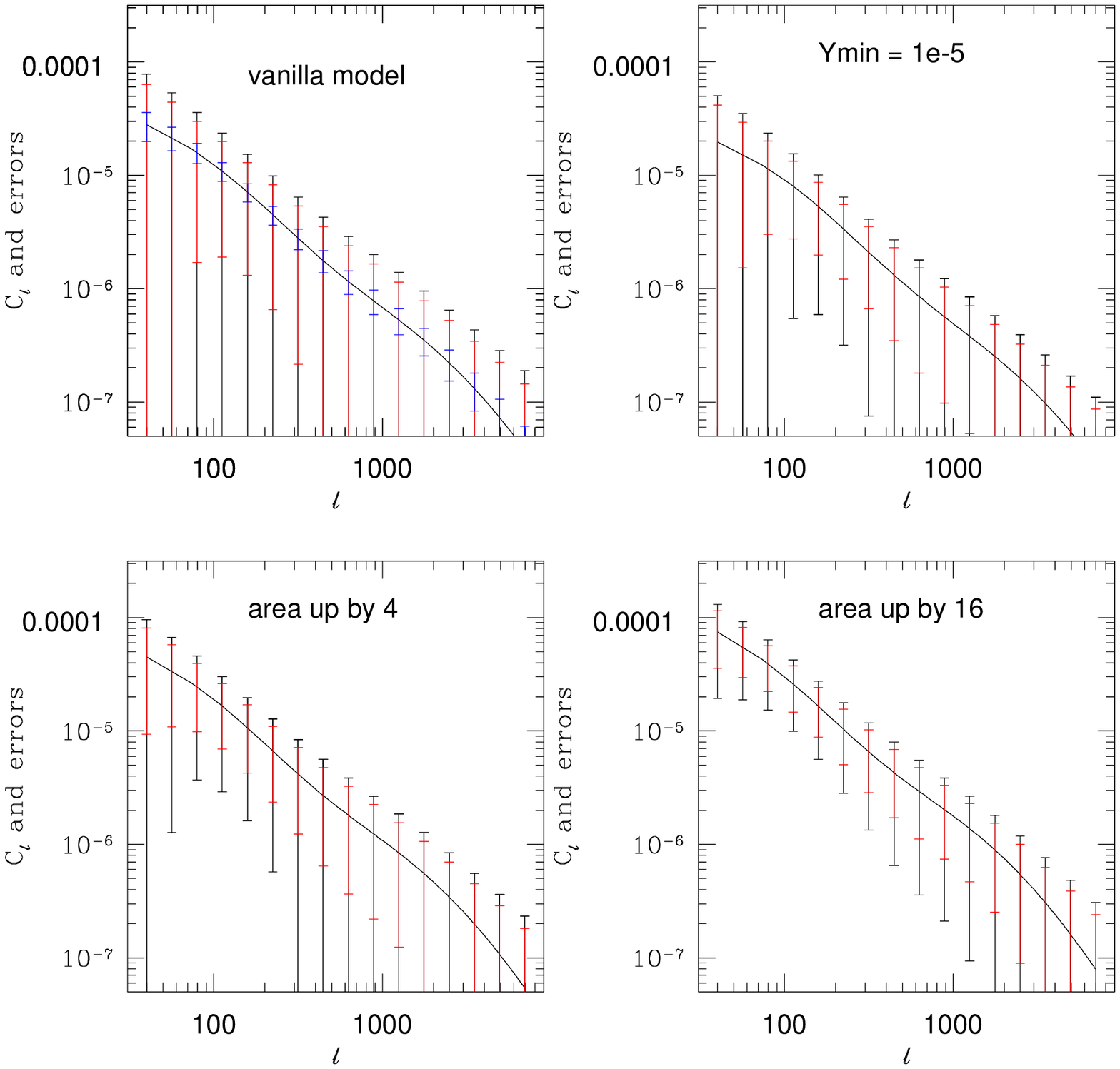}
\end{center}
\caption{The angular power spectrum with errors
from shot noise, sampling and cosmic variance error.
Error bars are for 100, 200 and 4000 square degrees for the
top two plots (vanilla model at left, $Y_{min} = 10^{-5}$ at
right), with the largest error bars corresponding to the
smallest area.
The lower figures correspond to the vanilla model except for the
change $Y_{min} \rightarrow 2 Y_{min,vanilla}$ ($4Y_{min,vanilla}$)
on the left (right), with the area going up by a factor
of 4 (16), relative to the vanilla model at upper left.  This naively
keeps the observing time fixed.  The error bars are equally spaced
in log $\ell$.
See the text for more details.
}
\end{figure}

In comparison, those for APEX and SPT are shown in figure 11.
The vanilla model has the same parameters as earlier (including
$Y_{min} = 1.7 \times 10^{-5}$), and error bars
are shown for 100, 200, and 4000 square degrees, representative
areas for data sets expected from APEX and SPT.  The largest
error bars are for the smallest area.
The plot on the upper right has the same observing area, but with
a minimum $Y_{min}$ value of $10^{-5}$.
This might occur if e.g. $T_{*}^{SZ}$ were 2.0 keV rather than 1.2,
a reasonable possibility given X-ray measurements and line of sight
contamination effects.  The sensitivity
is directly proportional to $Y_{min}$, so ``in principle'' an
experiment can fix $area/Y_{min}^2 $ for a given observing time,
producing a tradeoff between wide and shallow or narrow and deep.
The errors given in equation \ref{skycoverage} are shown for the vanilla
model correlation function and then compared to other possibilities
with fixed observing time.  Note this ignores how the efficiency 
and difficulty of cluster identification changes with $Y_{min}$.  The
errors also neglect the Poisson nature of the shot noise and the 3 and 4
point functions of the clusters, thus we are calling them ``naive''.
The bottom two graphs are the vanilla model plotted with a smaller
$Y_{min}$ with error bars
corresponding to the 100 and 200 square degree surveys with area up by
the factor of 4 or 16 for the left and right graphs respectively.
Note that as the sensitivity goes down that the power spectrum and
the shot noise go up, as mentioned earlier--only rarer and thus
more clustered objects are included.  The bin size is the spacing
between the error bars, equal in  log $\ell$.  However, the decrease in
error bars due to increased sky coverage outweighs the increase in shot
noise: shallow and wide appears naively preferable to deep and 
narrow.\footnote{Modifications of this to include Poisson shot noise
(Cohn \cite{Coh05}) appear to reduce the benefits of the shallower
surveys, but the exact comparison will require more assumptions about
which quantity is of interest, as well as inclusion of 3 and 4 point 
functions and an estimate of when the sampling goes from Poisson to
sub-Poisson (Casas-Miranda et al\cite{Casetal02}, Sheth \& Lemson 
\cite{SheLem99}).}

In fact, poorer sensitivity is not necessarily a bad thing, as the
omitted clusters have low mass and thus the strongest dependence on
the poorly understood gas physics. 
Exactly how deep is the most profitable will depend very
strongly on the normalization of the $Y(M)$ relation.  In
addition, the corresponding changes in the purity
and completeness of the surveys have not been taken into account
in this scaling argument,
how these properties scale will depend upon the cluster finding algorithms
in use as well.
Fixing the sensitivity and increasing the area, as SPT will do, is of course
a major improvement in any case.

In addition, even though the error bars are quite large, the measurement
contains useful information.  Theory gives smooth and 
known functions of $\ell$, $C_\ell$, so we can
bin $N_\ell$ nearby values of $\ell$, reducing the error.  For example,
we can use the combined shape of the power spectrum (20 bins equally
spaced in log $\ell$) and the 
errors in equation (30) to find one sigma contours in the 
$\sigma_8-\Omega_m$ plane.  This
is shown in figure 12 for marginalizing over
a 10\% prior and a 30\% prior on $T_*^{SZ}$ and
holding all the other uncertainties fixed for our ``vanilla'' model.
\begin{figure}
\begin{center}
\plotone{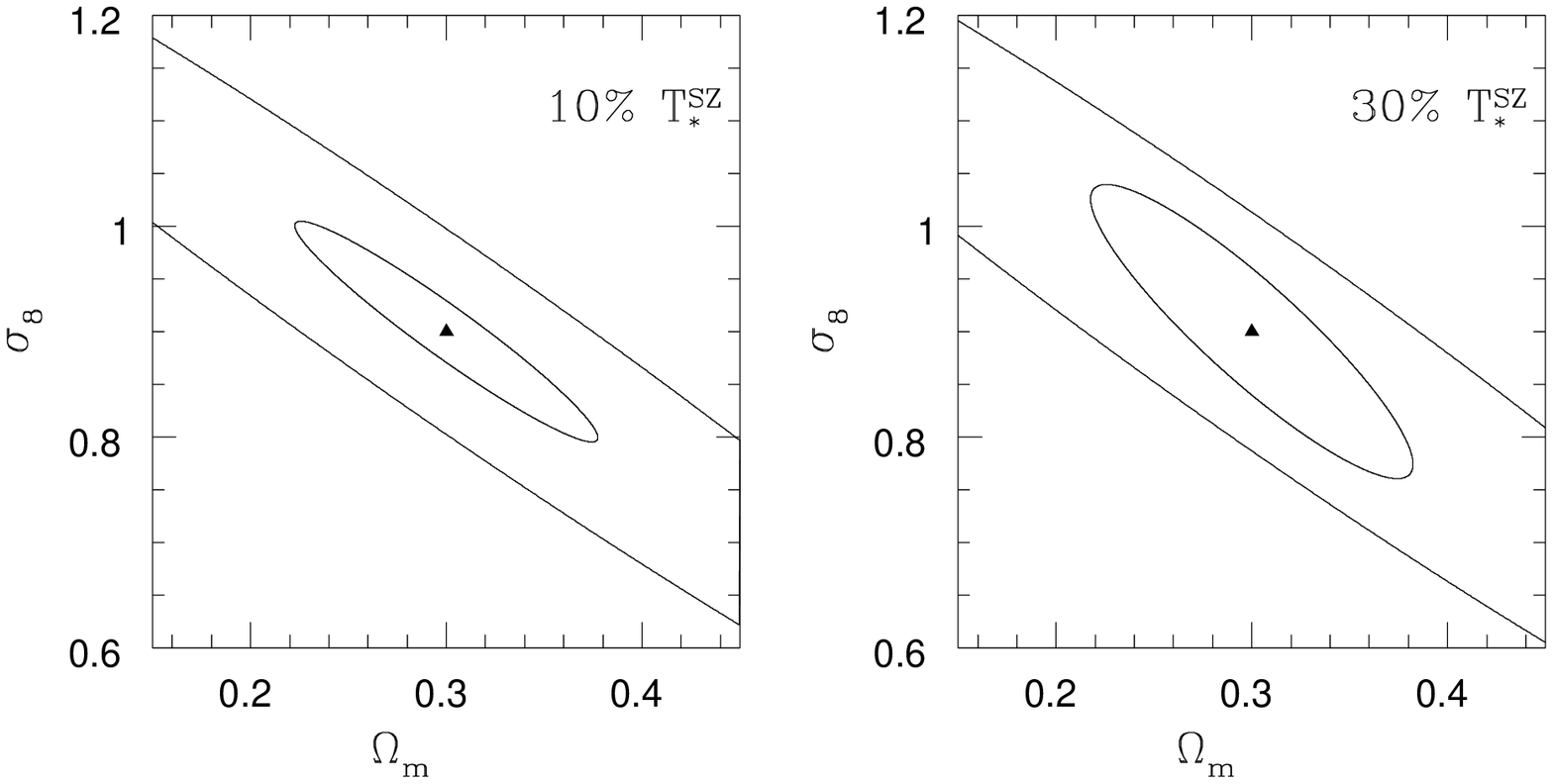}
\end{center}
\caption{The one sigma contours using the full power spectrum and
errors in equation (30), varying $\Omega_m$ and 
$\sigma_8$, marginalized with a 10\% prior on $T_*^{SZ}$ (left) and a
30\% prior on $T_*^{SZ}$ (right). The larger contour is for 200 square 
degrees (APEX) and the smaller one is for 4000 square degrees
(SPT). 
The true model is our ``vanilla'' model, shown by the point at the center. 
}
\end{figure} 
These error ellipses clearly illustrate the degeneracies inherent
in the effects of changing these two cosmological parameters for the angular 
power
spectrum.\footnote{The more precise full Poisson errors (Cohn \cite{Coh05})
will enlarge these error ellipses by 25-37 \% along the long axis and
by at most 5\% along the short axis, and rotate them by a small amount.  This 
more precise calculation still has the other assumptions used throughout
this paper however:
perfect cluster finding and no inclusion of the (unknown) 
cluster three and four point function contributions to the errors.  It does
include the (negligible) 10\% scatter in $Y(M)$.}
The closest error analysis to ours was done by Mei and 
Bartlett \cite{MeiBar04} who used the measurement of $w(\theta)$ at 30 arcminutes
and counts at the flux limit to get error contours for 
$\sigma_8$ and $\Omega_m$ of comparable size.   
Here instead we use the full power spectrum because our
intent is to illustrate the degeneracies in using the power
spectrum to constrain these parameters.
As Mei and Bartlett's degeneracy line in $\Omega_m, \sigma_8$ for $w(30')$ 
differs from the ellipse axes in figure 12, the two sets of constraints are 
complementary.

\section{Conclusions}
Forthcoming SZ surveys such as APEX are expected to observe one or two
orders of magnitude more clusters than currently in hand.  The 
angular power spectrum will be an immediate result once clusters
have been identified. There still exists uncertainty in theoretical 
predictions:  
we examined how those uncertainties affect the cluster power spectrum.

We calculated the angular power spectrum for
different reasonable mass functions, biases, 
mass-temperature normalizations $T_*^{SZ}$
and gas fractions $f_{gas}$ and found these changes are comparable to
changes due to cosmological parameters of interest such as $\sigma_8$ within
current ranges of interest.  In particular we identified which modeling 
uncertainties mimicked changes in the cosmological parameters 
considered (by finding the relevant cosmological parameter values) 
and which did not.  Some of these modeling and cosmological dependencies
have been studied for the correlation function in two dimensions or
the three dimensional power spectrum.  Different subsets with different
fixed assumptions have been considered in previous literature at different
times.  By combining all these variations in
a homogeneous manner, and including several more which have been identified
since, the relative importance of the various modeling 
assumptions can be more directly assessed.

We find that progress on several fronts is needed before the scientific
harvest from these experiments can be fully realized.

The uncertainties in the mass function of clusters 
and the bias can both be improved with simulations.  We showed the 
differences between many commonly used ones are significant in 
comparison to the uncertainties for the cosmological parameters of interest.

We have summarized much of the observational and theoretical work
on the $Y(M)$ relation, important uncertainties still remain.  
The normalization of the $Y(M)$ relation
needs considerable observational and/or simulation input 
(this point has also been emphasized previously by other authors 
as referenced in the text).  Current experiments such as SZA might
be able to do this calibration when combined with
other measurements for mass, as long as the normalization is not
redshift dependent.
Simulations with gas are known to have incomplete treatments of
physics but can be used as a guide, e.g. to calibrate the effect of 
projection on the mass temperature relation, which gives a systematic
increase of the $Y(M)$ normalization of about 8\%. 
We included scatter in the $Y(M)$ relation, however this was a small
effect:  the angular power spectrum changed by less than 1\%
when the $Y(M)$ relation was taken to have a scatter (as recent 
hydrodynamic simulations suggest) of 10\%.  The effects of mergers
seem to be small, once the normalization is fixed; we argue
that the effect of most of the parameters is just to decide whether
objects are included in the survey or not, and mergers have the
largest effect on the largest clusters, which tend to be included
anyhow for the low values of $Y_{min}$ under consideration here.

We also studied the experimental uncertainties which exist even 
in the absence of systematic uncertainty. We compared area to
sensitivity $Y_{min}$ and found that naively a shallower wider
survey is more powerful.

We focused on the angular power spectrum alone, though
of course complementarity is key to progress.  Complementary
data will even be provided from survey producing the
power spectrum itself.  
A survey providing an angular power spectrum will also produce
number counts per square degree and $dN/dY$, number
counts as a function of $Y$.  In addition, the temperature correlation
function and perhaps $dN/dz$ will be available.  Various combinations
of these quantities have been analyzed in the literature.  Mei
and Bartlett considered number counts and the angular correlation function,
and combined it with the number counts from X-ray, for instance.
This is one example of self-calibration (Levine, Schulz \& White
\cite{LevSchWhi02}); adding the angular power spectrum to other measurements
will increase the leverage of all of them.

J.D.C. thanks G. Evrard, A. Kravtsov, E. Reese and W. Hu for discussions
and S. Mei for help with comparing with her work;
K.K. thanks P. Zhang, W. Hu and J. Weller for useful discussions. 
We both especially thank M. White for numerous discussions, early
collaboration on this project and comments on the draft, and the
anonymous referee for many constructive suggestions.
This work was supported in part by DOE, NSF grant NSF-AST-0205935 and 
NASA grant NAG5-10842.

\end{document}